\documentclass[preprint,eqsecnum,aps,keywords,nofootinbib]{revtex4}
\usepackage{graphicx,amssymb,amsmath}
\usepackage{epsfig}
\usepackage{float}
\usepackage{latexsym}
\usepackage{rotating}

\begin{document}
\title[Geodesic motion in R -charged black hole space-times]{Geodesic motion in R- charged black hole space-times}

\author{Rashmi Uniyal}
\affiliation{Department of Physics, Gurukul Kangri Vishwavidyalaya,
Haridwar, Uttarakhand 249 404, India}
\email{rashmiuniyal001@gmail.com $\&$ hnandan@iucaa.ernet.in}
\author{Anindya Biswas}
\affiliation{Department of Physics, Ranaghat College, Ranaghat, West Bengal 741 201, India}
\email{ani$\_$imsc@yahoo.co.in}
\author{Hemwati Nandan}
\affiliation{Department of Physics, Gurukul Kangri Vishwavidyalaya,
Haridwar, Uttarakhand 249 404, India}
\email{hnandan@iucaa.ernet.in}
\author{K. D. Purohit}
\affiliation{Department of Physics, Hemwati Nandan Bahuguna Garhwal University,
Srinagar Garhwal, Uttarakhand 246 174, India}
\email{kdpurohit@rediffmail.com}

\begin{abstract}
We study the geodesic motion of massive and massless test particles in
the background of a particular class of multiple charge black holes in gauged
supergravity theories in $D=4$. We have analysed the horizon structure
along with the nature of the effective potentials for the case of four
equal charges. In view of the corresponding effective potentials, we
have  discussed all the possible orbits in detail for different values
of energy and angular momentum of the incoming test particles. The periods
for one complete revolution of circular orbits and the advance of
perihelion of the planetary orbit have also been investigated in
greater detail for massive test particles. We have also discussed the
time period of unstable circular motion and cone of avoidance of
massless test particles in detail. All the corresponding results obtained for
massive and massless test particles are then compared accordingly.
\end{abstract}
\pacs{00.00, 20.00, 42.10}
\maketitle
\section{Introduction:}
\noindent
Over the last century, the \textit{Einstein's theory of gravitation} i.e. General Relativity (GR)\citep{Eins,Poi,Wald,Joshi} has been extremely successful to understand various observational facts like gravitational redshift, the precession of Mercury's orbit, the bending of light etc. On the other hand, various black hole spacetimes with or without rotation were obtained as exact solutions of Einstein field equations in GR namely Schwarzschild metric \citep{Schwar}. Other black hole spacetimes including charges and rotations have also been discovered. 
Though GR has enjoyed the great success, but it is still not a complete theory to understand the physics at sufficiently smaller length scale e.g. near the spacetime singularity which arises in case of gravitational collapse \citep{Joshi}.
In the vicinity of the spacetime singularity, the quantum effects should be taken care of seriously and recently string theory has become the promising candidate for the same purpose which comprises gravity in the frame work of quantum theory {\citep{Sen2005,Zwie2004,Pol1998}}.
\\
\noindent
In continuation pursuit of the search for a quantum theory of gravity, the gauged supergravity theories{ \citep{Wit1982,Nic1982,Duff1986}} have also captured considerable attention in recent times. 
In such models of gravity, the maximally supersymmetric gauged supergravity is realized as truncation of string theory or $D = 11$ dimensional M theory compactified on a sphere where the gauge group is the isometry group of the sphere. In particular, in $AdS/CFT$ correspondance \citep{Malda1998}, this gauge group becomes the R-symmetry group of the boundary CFT which is related to the string theory on anti-de Sitter space.
Black hole solutions in the $AdS$ sector have drawn much attention to understand strongly coupled gauge theory at finite temperature on the CFT side \citep{Witt1998}, \citep{Witten1998}. For the present work, we consider four dimensional charged black holes in $N=8${\footnote {$N$ is the number of supermultiplates.}}, gauged supergravity upto four charges. 
The construction of these black holes have been explicitly performed in Ref. \citep{Liu1999} and according to \citep{Cvetic1999}, such charged black hole conﬁgurations are termed as R-charged black holes.
In gauged $N=8$ supergravity models, the bosonic part of the complete Lagrangian has a negative cosmological constant $\Lambda$ proportional to the square of the gauge coupling constant{\footnote {$\Lambda \sim -{g^2}$, where $g \equiv {1\over l}$ and $l$ is the $AdS$ length scale.}} $g$ \citep{Duf1999} and the black hole solutions are asymptotically $AdS$.\\
\noindent
Motivated from the study of motion of massless and massive test particles in the
background of various black holes in GR \citep{Hagi1931,Kagra2010,Grunau2011,Kagra2011,Heck2008,Hec2008,Heckm2008,Heck2010,Enol2011,Grunau2012,
Grunau2013,Oliv2011,Vill2013,Sor2015,Uniyal2015,Ghosh2010,Nandan2010,Das2012} and other alternative theories of gravity like string theory \citep{Koley2003,Kuniyal2015,Uniyal2014,Das2009,Gad2010,Fernando2012,Oliva2013,Saskia2013}, we try to make an attempt to understand the geodesics of massive neutral as well as massless particles in the exterior region of such R-charged black holes.
Such investigations may indeed useful to capture the effect of cosmological constant in addition to the charges on the motion of the test particles.
The main objective of this paper is to study all the possible orbits of test particles in these class of spacetimes.
We analyze the equivalent one dimensional effective potentials in order to study the geodesics equations of motion of the test particles in R-charged black holes background. On the other hand, we also focus our study on the various parameters (mass parameter, multiple charge parameters and gauge coupling constant) of the solutions and present various possible orbits by tuning those parameters. Due to the presence of four non zero charges, the form of the one dimensional effective potentials are little complicated in our case. 
We mainly concentrate on the numerical approach without performing the analytic solutions \citep{Heck2008,Hec2008,Heckm2008}. The form of potentials can be retraced back to the known form in the limit of zero charges which otherwise takes quite nontrivial form.\\
\noindent
This paper is organised as follows. In the next section, we discuss about the spacetime of our interest and the horizon structure of these particular class of R-charged black holes 
\citep{Liu1999} with multiple charges, whereas the ungauged situation (i.e. $g=0$) is same as the Schwarzschild black hole with flat asymptotes in GR. In section III, we derive the geodesics equation and obtain the form of the effective potential. In section IV, we study geodesics equation of motion for massive neutral particles (i.e. the timelike geodesics) under one dimensional effective potential by considering different values of the parameters involved in. The radial and non-radial geodesics are studied analytically and otherwise numerically. 
Using the effective potential techniques, the motion of test particles and the structure of corresponding orbits are discussed in greater detail. In order to have a complete analysis of the geodesic motion in the background of the abovementioned black hole spacetime, we also study the motion of massless test particles (i.e. null geodesics) in section V. Finally, we conclude and summarize our results.
\section{The spacetime with multiple charges}
\begin{figure}[h!]
\centerline{\includegraphics[scale=0.3]{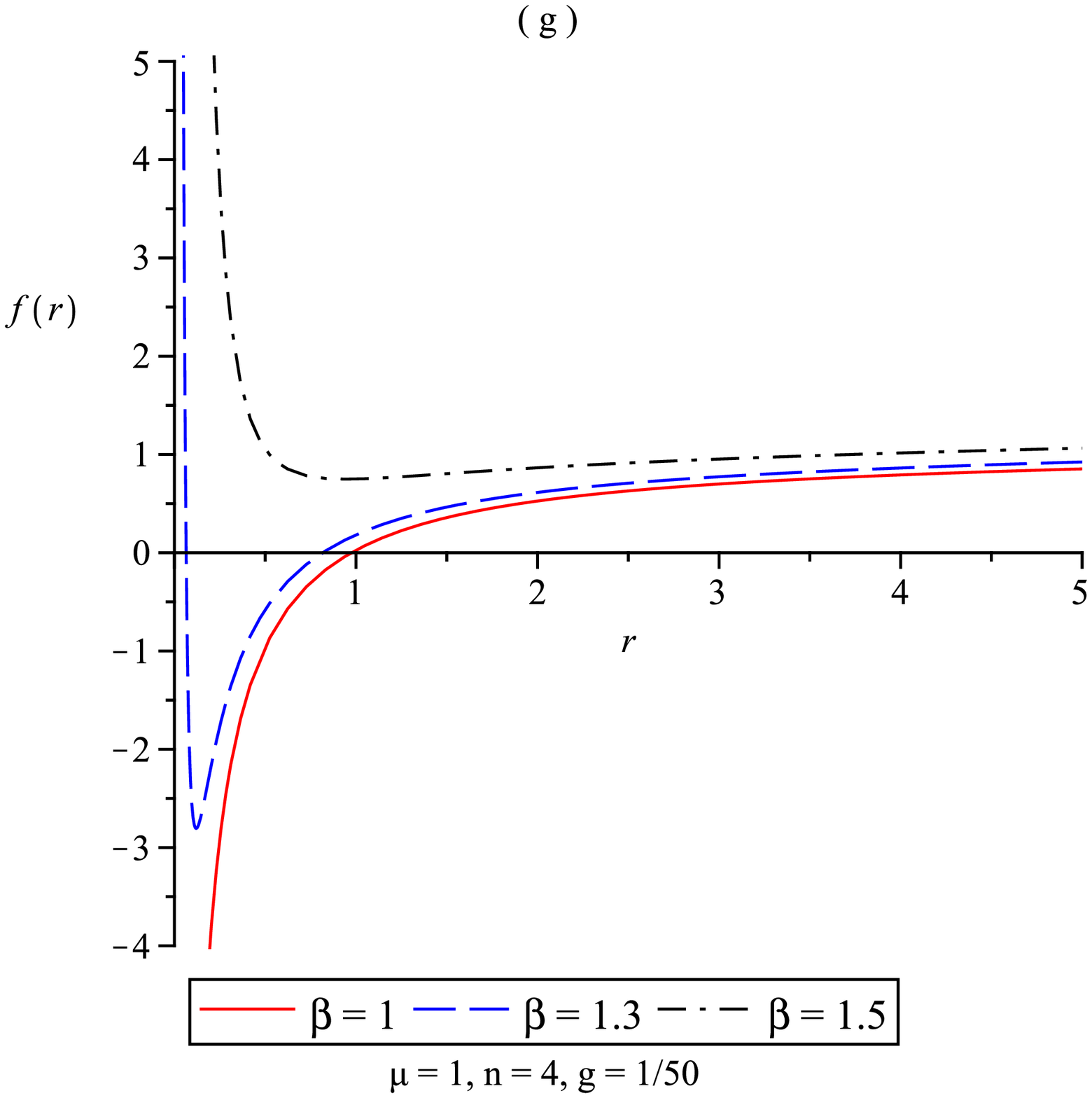}
\hskip1mm \includegraphics[scale=0.3]{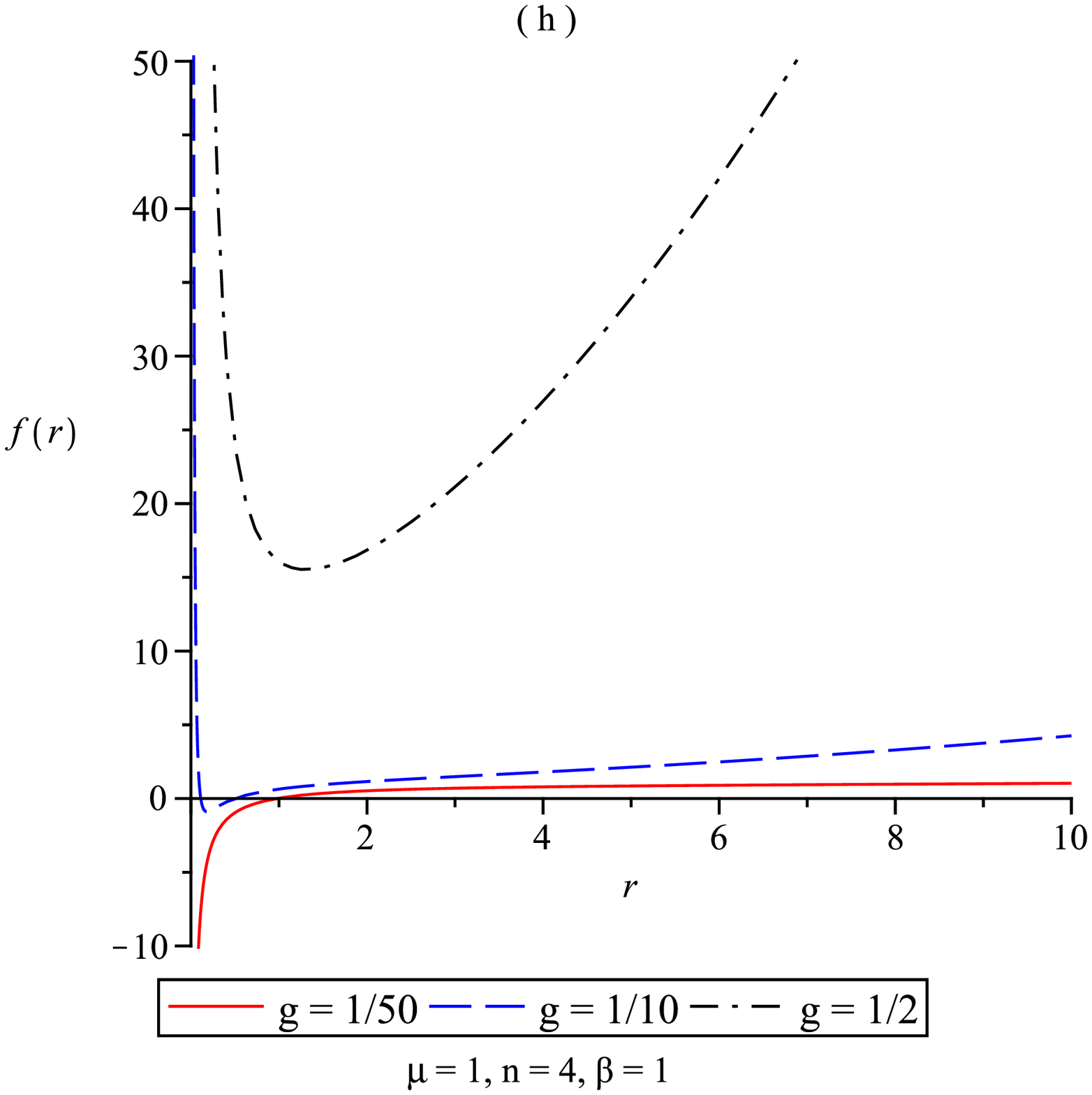}}
\caption{\small{The lapse function $f(r)$ for R-charged black hole: (a) and (b) represent the single charge case, (c) and (d) corresponds to two charge case, (e) and (f)  are for three charge. Further (g) and (h) represent the case corresponds to four equal charges.
\protect\label{horizon_1_4}}}
\end{figure}

\noindent
We consider the following spacetime metric of four electric charge black hole solution in $D=4$, $N=8$ gauged supergravity,
\begin{equation}
ds^2 = - (H_1H_2H_3H_4)^{-1/2}fdt^2 + (H_1H_2H_3H_4)^{1/2}(f^{-1}dr^2+r^2d\Omega^2_{2,k}),
\label{met1}
\end{equation}
where
\begin{equation}
H_\alpha = 1+\frac{\mu\sinh^2{\beta_\alpha}}{r}, ~~~~~~~~~~~
f= k-\frac{\mu}{r}+2g^2 r^2 (H_1H_2H_3H_4).
\label{eq:f_n_H}
\end{equation}
Here $g$ is the gauge coupling and $\mu$ is the non-extremal parameter like mass term in pure Schwarzschild black hole, 
$k$ can take three values i.e. $1$, $0$ and $-1$,
however in the present study, we consider only the case of black hole solutions with $k$ = 1.
The  $\beta_\alpha(\alpha=1......4)$ in eqs. (\ref{met1}) and (\ref{eq:f_n_H}) parametrizes the four electric charges. \\\
The scalar curvature $R$ for the above metric can be given as below,
\begin{equation}
R = {1\over{\mathcal H^{1/2}}}\Big[{{2\over r^2} (1-f)} -{4f^\prime\over r}-{2f\over r}
({\mathcal H^\prime\over \mathcal H})+{3\over 8}f({{{\mathcal H'}^2}\over \mathcal H^2})-{f\over 2}({\mathcal H^{''}\over \mathcal H})\Big],
\label{Ricci}
\end{equation}
where $\mathcal H=H_1H_2H_3H_4$.
One can easily verify that for $H_\alpha=1$ (no charge solution), the scalar curvature $R=-8g^2$, which is exactly same as that of an Anti-de Sitter Schwarzschild black hole. \\
Let us first discuss about the event horizons of the metric eq.(\ref{met1}) in presence of 
$n$ equal charges. So for $n$ equal charges with $k$ = 1, equation (\ref{eq:f_n_H}) reduces to,
\begin{equation}
H = 1+\frac{p}{r}, ~~~~~~~~~~~
f= 1-\frac{\mu}{r}+2g^2 r^2 H^n,
\label{eq:f_n_equal_q}
\end{equation}
with $p={\mu}{\sinh^2}\beta$ and $n=0,1,2,3,4$. 
The lapse function $f$ vanishes at the zeros of the equation given as,
\begin{equation}
r^{n-3}(r-\mu)+2{g^2}(r+p)^n=0.
\label{eq:f_r}
\end{equation}
The real and positive zeroes of eq.(\ref{eq:f_r}) are known as the horizons for the corresponding spacetime. Hence if a spacetime represents black hole $f$ should vanish at some positive value of $r$. 
For $n$ = 1, 2, 3, there is always a cubic equation for $r$ while for $n$ = 4, we have a quartic equation for $r$.
The roots of eq.(\ref{eq:f_r}) in a closed form are really lengthy and complicated, so
 are not discussed over here, rather we present the graphical presentation of the
position of the horizons by imposing certain conditions among the parameters of the black holes. In presence of one and two equal charges one will always have a horizon as shown in the fig(\ref{horizon_1_4}$a, b, c, d$) for various values of $\beta$ and $g$. While in presence of three and four equal charges, there are some constraints on the parameters involved in the metric to represent a black hole with at least one horizon. In presence of three equal charges, the constraint on parameters for the presence of horizon is given by,
\begin{equation}
2{g^2}{{\mu}^2}{{\sinh}^6{\beta}}<{1}.
\label{eq:condition_bh_3q}
\end{equation}
The three type of plots in fig(\ref{horizon_1_4}$e$) or (\ref{horizon_1_4}$f$) clearly explain the facts that black curve does not satisfy the bound eq.(\ref{eq:condition_bh_3q}) with the values of parameters chosen. So it is obvious that there is a possibility of naked singularity in case of three equal charges if one crosses the bounds given in eq.(\ref{eq:condition_bh_3q}). In case of four equal charges present, the condition of getting at least one horizon is given by,
\begin{equation}
A+B-{\frac {1}{256}}<0,
\label{eq:condition_bh_4q}
\end{equation}
where 
$A={\sinh^{6}\beta}{g}^{4} \left( {\sinh^{2}\beta}+1 \right) ^{3}{\mu}^{4}$,\\
and $B=\frac{1}{32}\,{g}^{2} {\mu}^{2}
 \left( {\sinh^{8}\beta}+2\,{\sinh^{6}\beta}-\frac{7{\sinh^{4}\beta}}{2}-\frac{9{\sinh^{2}\beta}}{2}-{\frac {27}{16}}\right)$.\\
The zeros of the function $f(r)$ are presented in the plots of fig(\ref{horizon_1_4}$g, h$). In fig(\ref{horizon_1_4}$g$), one can see that for sufficiently small values of the constant $g$ the number of zeros for $f(r)$ changes. For certain values of $\beta$, one will get two zeros of the function $f(r)$ whereas smaller than that two zeros merge into a single zero and larger than that real zeros will disappear completely by leaving with a naked singularity as explained in \citep{Romans1992}. The same arguments are also plausible when the constant $g$ varies keeping charge parameters $\beta$ fixed, as shown in fig(\ref{horizon_1_4}$h$).    
\begin{figure}[h]
 \centerline{\includegraphics[scale=0.3]{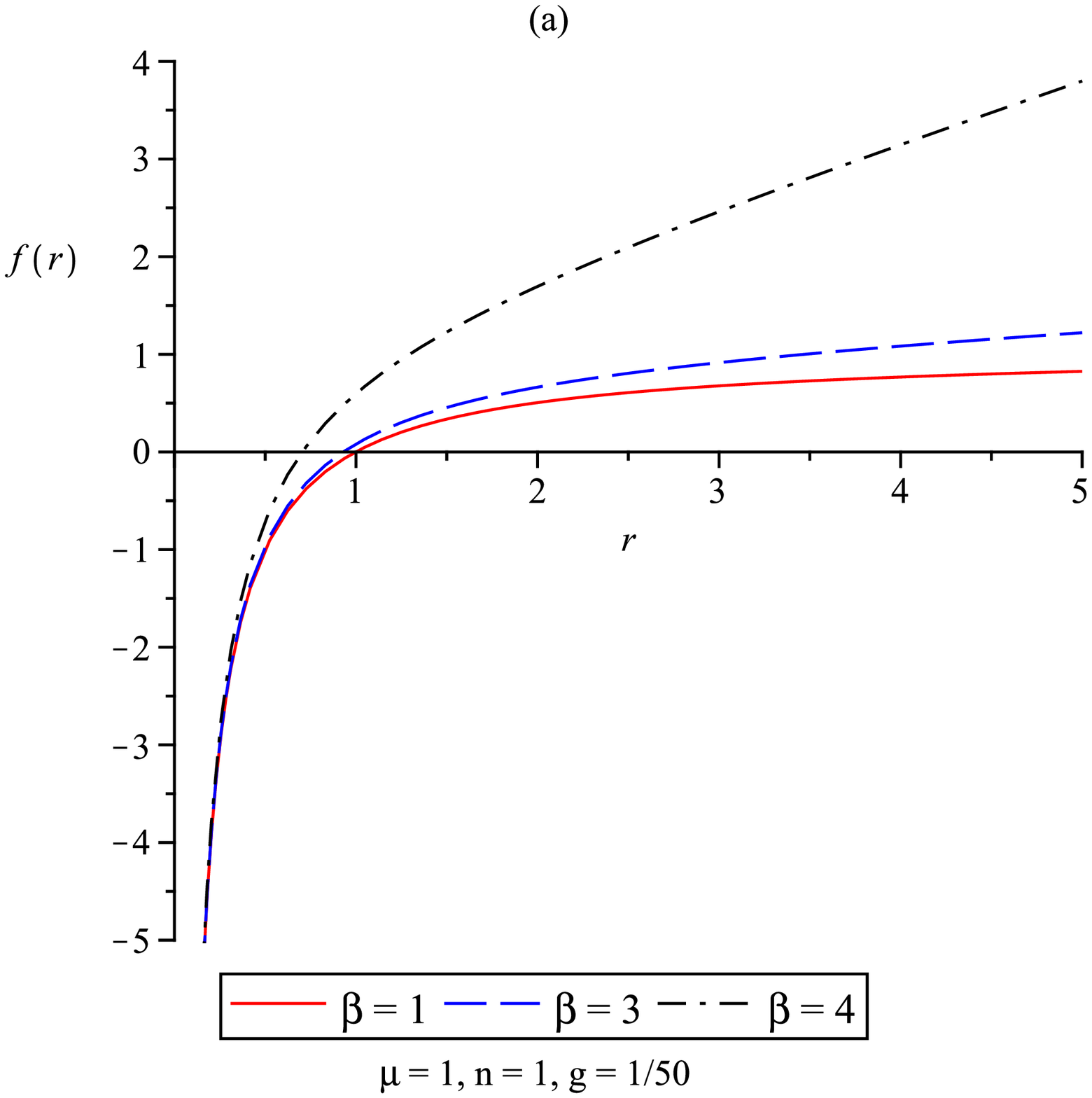}
 \hskip1mm \includegraphics[scale=0.3]{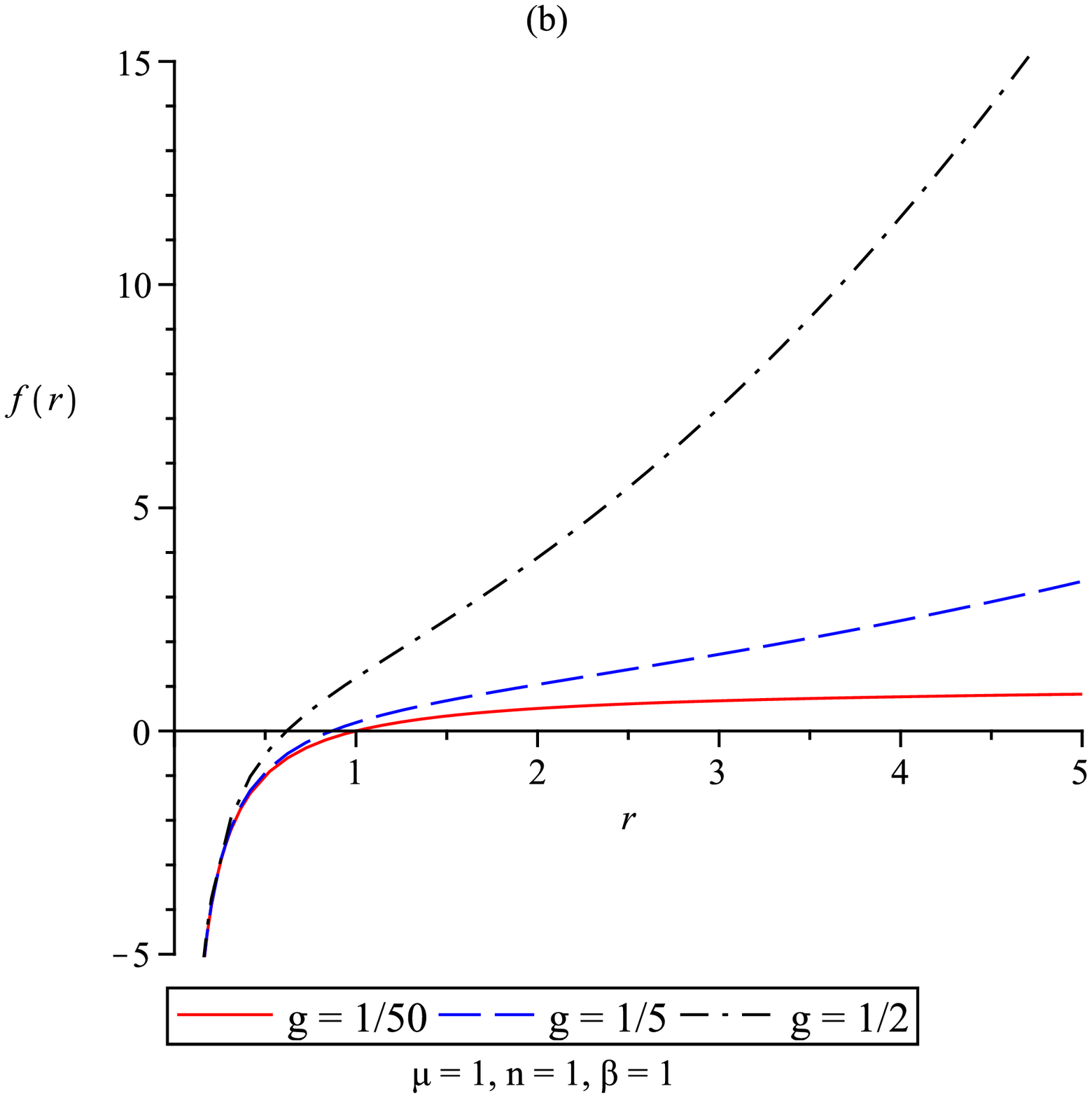}
 \hskip1mm \includegraphics[scale=0.3]{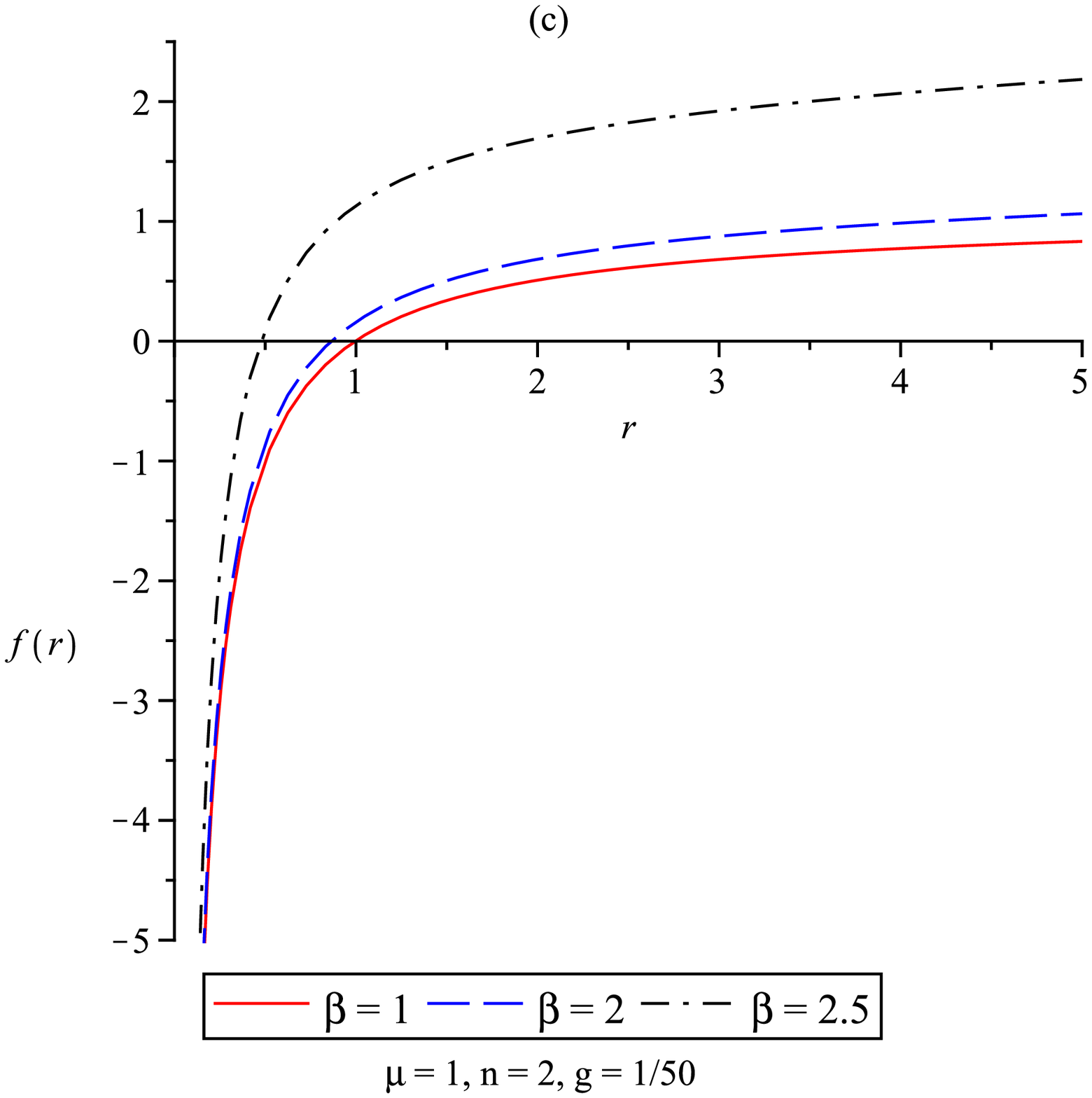}}
 \centerline{\includegraphics[scale=0.3]{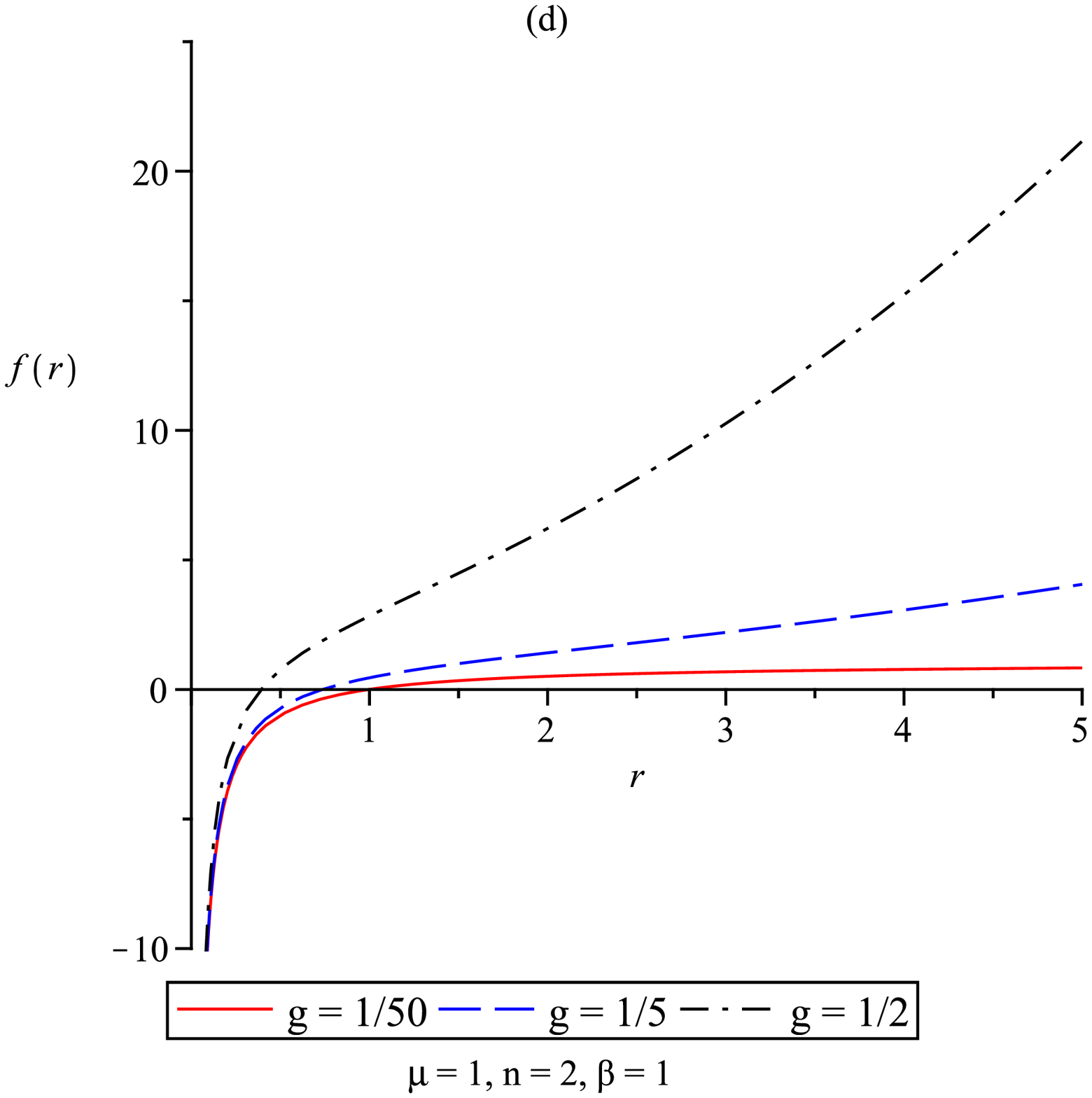}
\hskip1mm \includegraphics[scale=0.3]{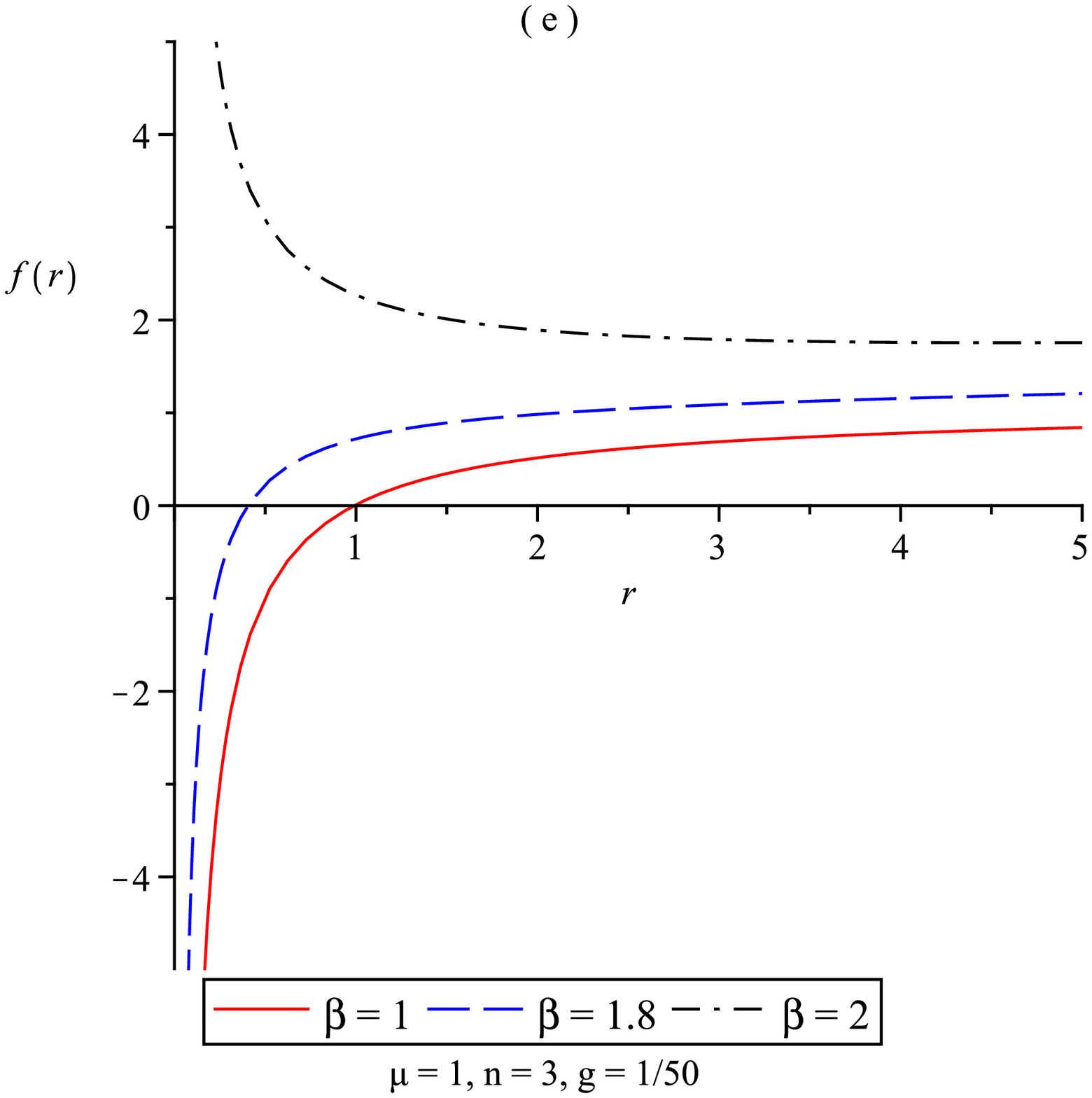}
\hskip1mm \includegraphics[scale=0.3]{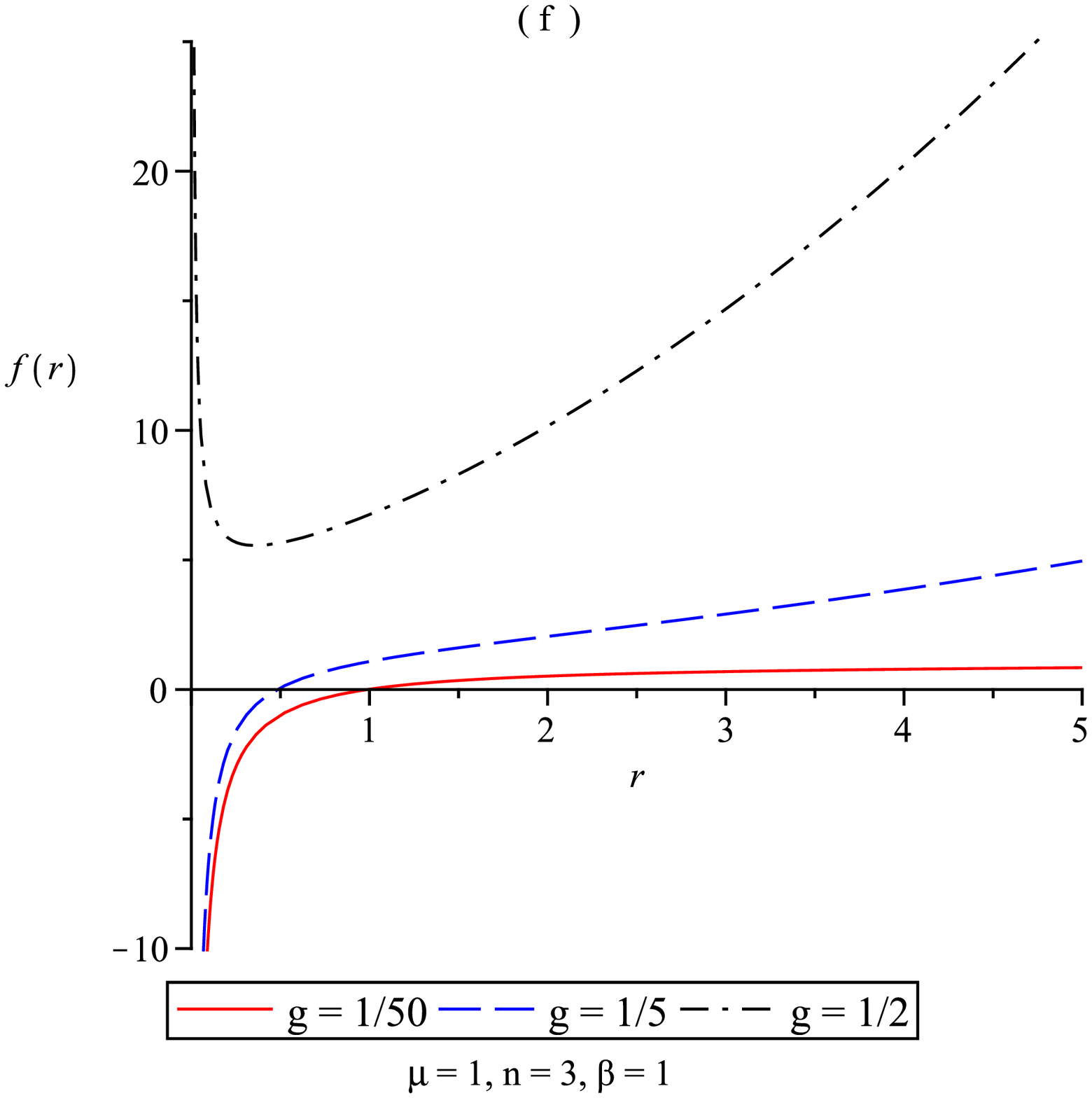}}
\nonumber
\end{figure}
\section{The Geodesics Equations and Effective Potential}
\noindent  
One can simplify metric given in eq.(\ref{met1}) by choosing $\theta={\pi\over2}$, to restrict the motion of the particles in the equatorial plane i.e. we restrict ourselves to study of the equatorial geodesics.\\
Using the following geodesic equation
\begin{equation}
{d^2x^\alpha\over{d\tau^2}} = -\Gamma^\alpha_{\beta\gamma}{{dx^\beta\over{d\tau}}{dx^\gamma\over{d\tau}}},
\label{Geo}
\end{equation}
we obtain
\begin{equation}
\ddot t + {1\over 2}\Big({f^\prime\over f}-{\mathcal H^\prime\over
{2\mathcal H}}\Big)\dot t \dot r=0,
\label{Time}
\end{equation}
\begin{equation}
\ddot r + {f^2\over{2\mathcal H}}\Big({f^\prime\over f}-{\mathcal H^\prime\over
{2\mathcal H}}\Big)\dot t^2-{1\over 2} \Big({f^\prime\over f}-{\mathcal H^\prime\over
{2\mathcal H}}\Big)\dot r^2-rf\Big(1+{r\over 4}{\mathcal H^\prime\over\mathcal H}\Big)\dot\phi^2=0,
\label{rr}
\end{equation}
\begin{equation}
\ddot\phi + \Big({\mathcal H^\prime\over 4\mathcal H}+{1\over r}\Big)\dot r\dot\phi=0,
\label{Phi}
\end{equation}
where dot denotes the derivative with respect to proper time $\tau$.\\
For a timelike geodesics $(u^\mu u_\mu=-1)$,
\begin{equation}
-\mathcal H^{-{1\over 2}} f \dot t^2 + \mathcal H^{1\over 2} f^{-1} \dot r^2 + \mathcal H^{1\over 2} r^2\dot \phi^2 = -1.
\label{timelike}
\end{equation}
From the constant of motion, the first integrals of eqs. (\ref{Time}) and (\ref{Phi}) result as,
\begin{equation}
\dot t = E \mathcal H^{1\over 2} f^{-1},
\label{Com1}
\end{equation}
\begin{equation}
\dot \phi = L \mathcal H^{-{1\over 2}} r^{-2},
\label{Com2}
\end{equation}
where $E$ and $L$ are the integration constants which represent the conserved energy and angular momentum of the test particles respectively.\\
Using eqs.(\ref{Com1}) and (\ref{Com2}), one can now rewrite the timelike 
constraint eq.(\ref{timelike}) in the following form,
\begin{equation}
{1\over 2}\dot r^2 + {1\over 2}\Big( \frac{L^2 f}{r^2 \mathcal H}+\frac{f}{\mathcal H^{1\over 2}}\Big) = \frac{E^2}{2}.
\label{Eom}
\end{equation}
The effective potential for timelike geodesics may be identified from eq.(\ref{Eom}) as,
\begin{equation}
V_{eff} = \frac{f}{2\mathcal H^{1\over 2}}
\Big( \frac{L^2}{r^2 \mathcal H^{1\over 2}}+1\Big).
\label{eq:Potential_Non_Radial}
\end{equation}
One can reproduce the effective potential for the schwarzschild black hole with $AdS$ asymptote as a limiting case when $H_\alpha=1$. 
It seems that under scaling of the metric components like $f\rightarrow f\mathcal H^{-{1\over 2}}$ and $r^2\rightarrow r^2\mathcal H^{1\over 2}$, one can exactly write down all the corresponding equations above for the case of a schwarzschild black hole in flat space \citep{Wald}.
However, the effective potential for null geodesics is given as,
\begin{equation}
V_{eff} = \frac{fL^2}{2r^2\mathcal H}.
\label{eq:Potential_Non_Radial_Null}
\end{equation}
The difference in the nature of effective potentials for both the cases as
given in eq.(\ref{eq:Potential_Non_Radial}) and eq.(\ref{eq:Potential_Non_Radial_Null}) manifests itself in the structure of
orbits as presented in the later sections.

\section{Nature of effective potential and classification of orbits for Timelike Geodesics}
\noindent From effective potential given in eq.(\ref{eq:Potential_Non_Radial}) for particles moving along the timelike geodesics, the equation of motion (\ref{Eom}) is investigated for two cases: radial and non-radial geodesics in the following subsections, respectively.    
\subsection{Effective potential for radial geodesics ($L=0$)}
\begin{figure}[h]
\centerline{\includegraphics[scale=0.3]{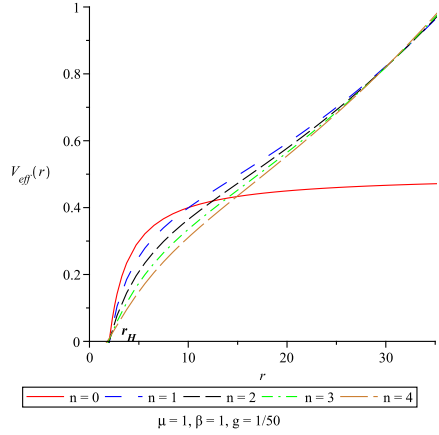}
}
\vspace*{4pt}
\caption{\small{Variation of effective potential with $r$ for radial geodesics with different values of $n$. \protect\label{fig:potential_01}}}
\end{figure}

\noindent For radial geodesic (with zero angular momentum $L$), the effective potential given in eq.(\ref{eq:Potential_Non_Radial}) reduces to the following form,
\begin{equation}
V_{eff} = \frac{f}{2\mathcal H^{1\over 2}}.
\label{eq:Potential_Radial}
\end{equation}
The behaviour of effective potential is shown in fig.(\ref{fig:potential_01}) for a particular set of parameters, considering all values of $n$ ranging from $0$ to $4$.
\par\noindent 
If the particle is released from rest at a distance $r={r_i}$, the initial energy of the particle is given as,
\begin{equation}
{E^2}\left({r_i}\right)=\frac{1-\frac{\mu}{r_i}+2{g^2}{{r_i}^2}\mathcal H\left({r_i}\right)}
{\sqrt{\mathcal H\left({r_i}\right)}}.
\label{eq:Energy_r_i}
\end{equation}
So from the radial equation of motion, one obtains
\begin{equation}
{\dot{r}}^2={E^2}\left({r_i}\right)-\frac{1-\frac{\mu}{r}+2{g^2}{{r}^2}\mathcal H\left({r}\right)}{\sqrt{\mathcal H\left({r}\right)}},
\label{eq:EoM_radial}
\end{equation}
which can be integrated as,
\begin{equation}
\mathcal{\tau}\left(r\right)={\Large{\int_{r_i}}^r}\frac{dr}{\sqrt{{E^2}\left({r_i}\right)-\frac{1-\frac{\mu}{r}+2{g^2}{{r}^2}\mathcal H\left({r}\right)}{\sqrt{\mathcal H\left({r}\right)}}}}.
\label{eq:tau_integral}
\end{equation}
\begin{figure}[h]
\centerline{\includegraphics[scale=0.3]{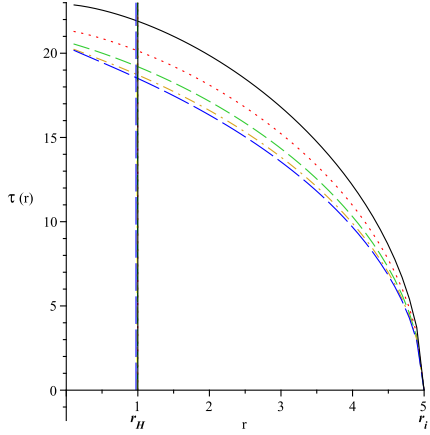}}
\vspace*{4pt}
\caption{\small{Proper time $\tau$ as a function of $r$ with $\mu=1$, $\beta=1$, $g=0.02$ and ${r_i}=5$. Here solid, dotted, dashed, dot dashed and long dashed lines represent $n=0,1,2,3,4$ cases respectively and $r_{_H}$ shows the position of event horizon. \protect\label{fig:proper_time_radial}}}
\end{figure}
In order to know the proper time experienced by a particle falling from $r_i$ to coordinate radius $r$, one needs to evaluate the integral given in eq.(\ref{eq:tau_integral}).
Since the result in a closedform is not obvious here, we look for numerical solutions.
The proper time given by eq.(\ref{eq:tau_integral}) is visually presented in fig.(\ref{fig:proper_time_radial}) which indicates that the particle falls towards the horizon in a finite proper time. 
It can also be observed from this figure that the proper time to reach horizon decreases with the increasing number of charges.
\subsection{Effective potential for non-radial geodesics ($L\neq 0$)}
\begin{figure}[h]
\centerline{ \includegraphics[scale=0.25]{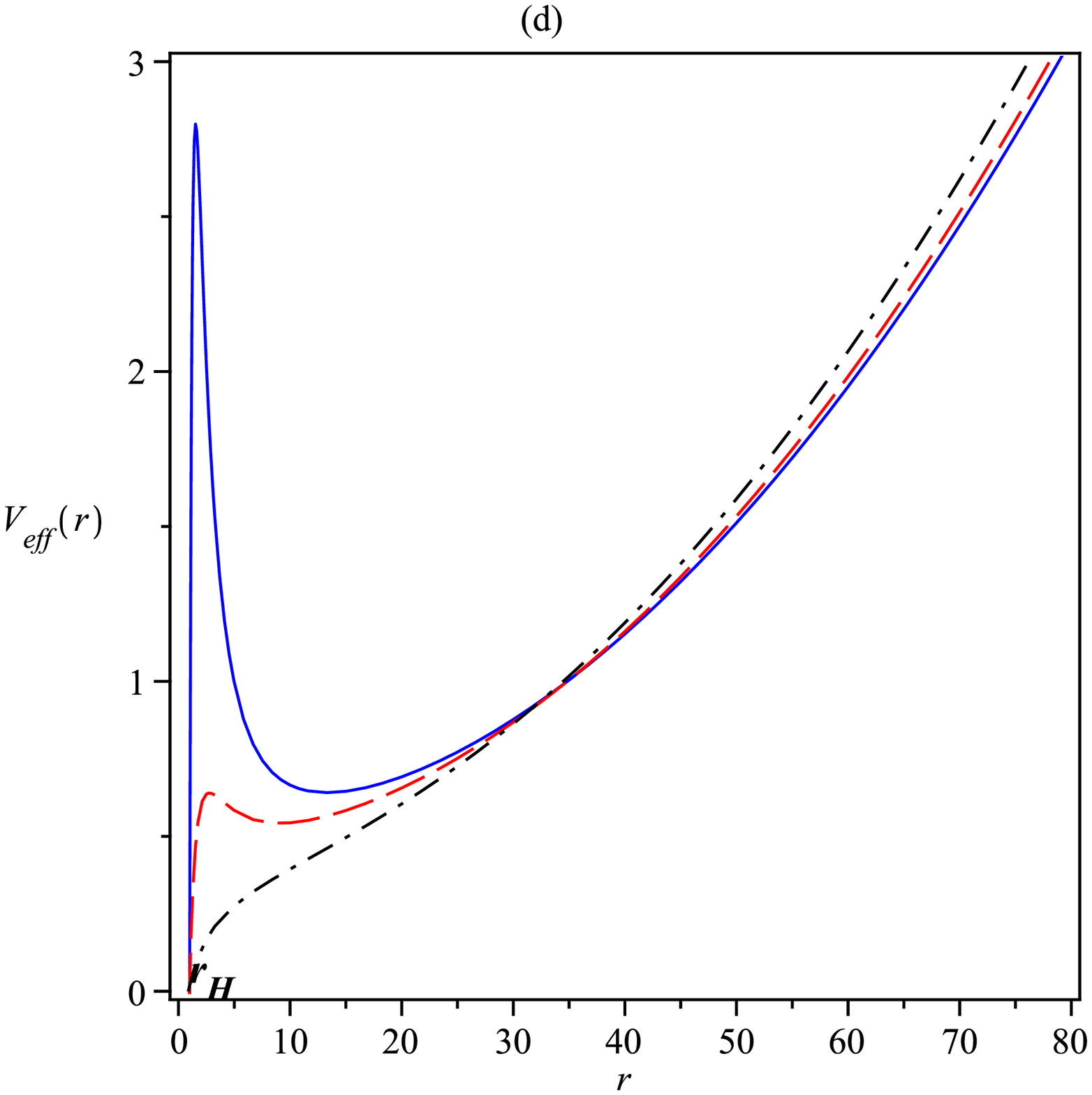}
\hskip1mm \includegraphics[scale=0.25]{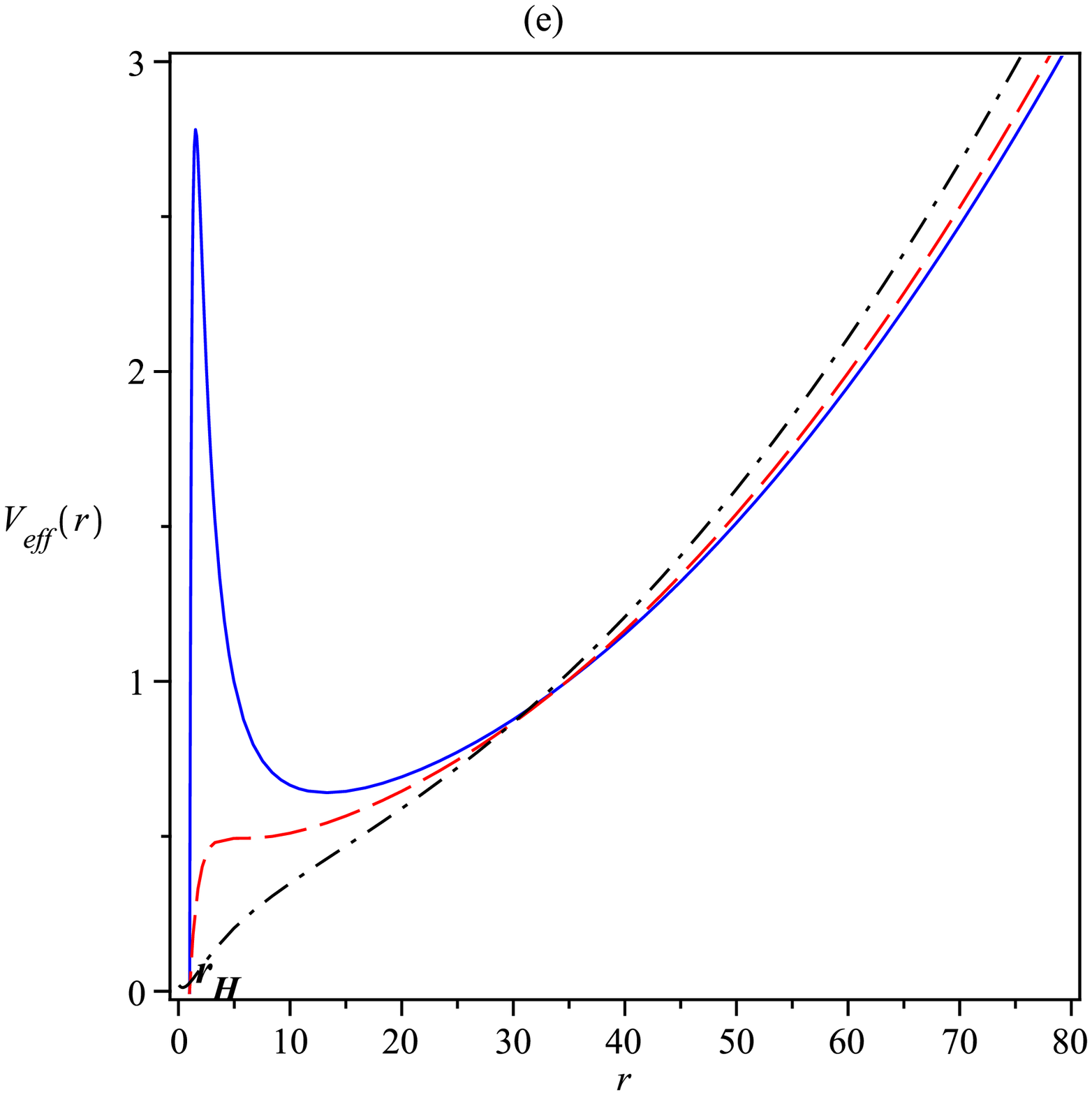}}
\caption{\small{Effective potential for non radial geodesic with $\mu=1$, $g=0.02$, $L=6$, $r_{_H}$ represents the position of event horizon. Fig.(a)-(e) represent the cases of $n=0,1,2,3,4$ respectively, where for solid line $\beta=0.1$, for dashed line $\beta=1$ and for dot-dashed line $\beta=1.5$ (while $n=0$ case does not contain any $\beta$ term). 
\protect\label{fig:potential_nr_01}}}
\end{figure}

\begin{figure}[h]
\centerline{ \includegraphics[scale=0.25]{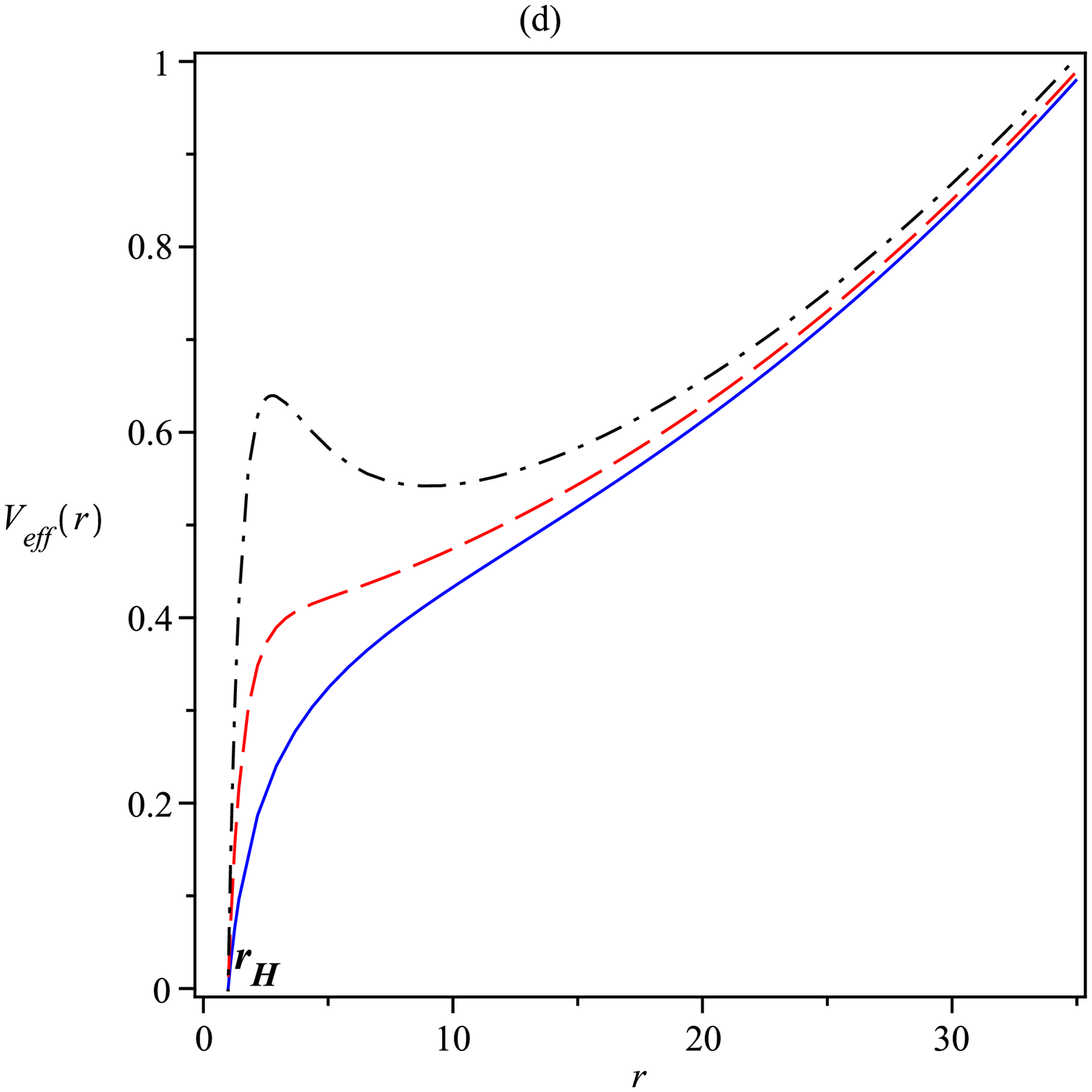}
\hskip1mm \includegraphics[scale=0.25]{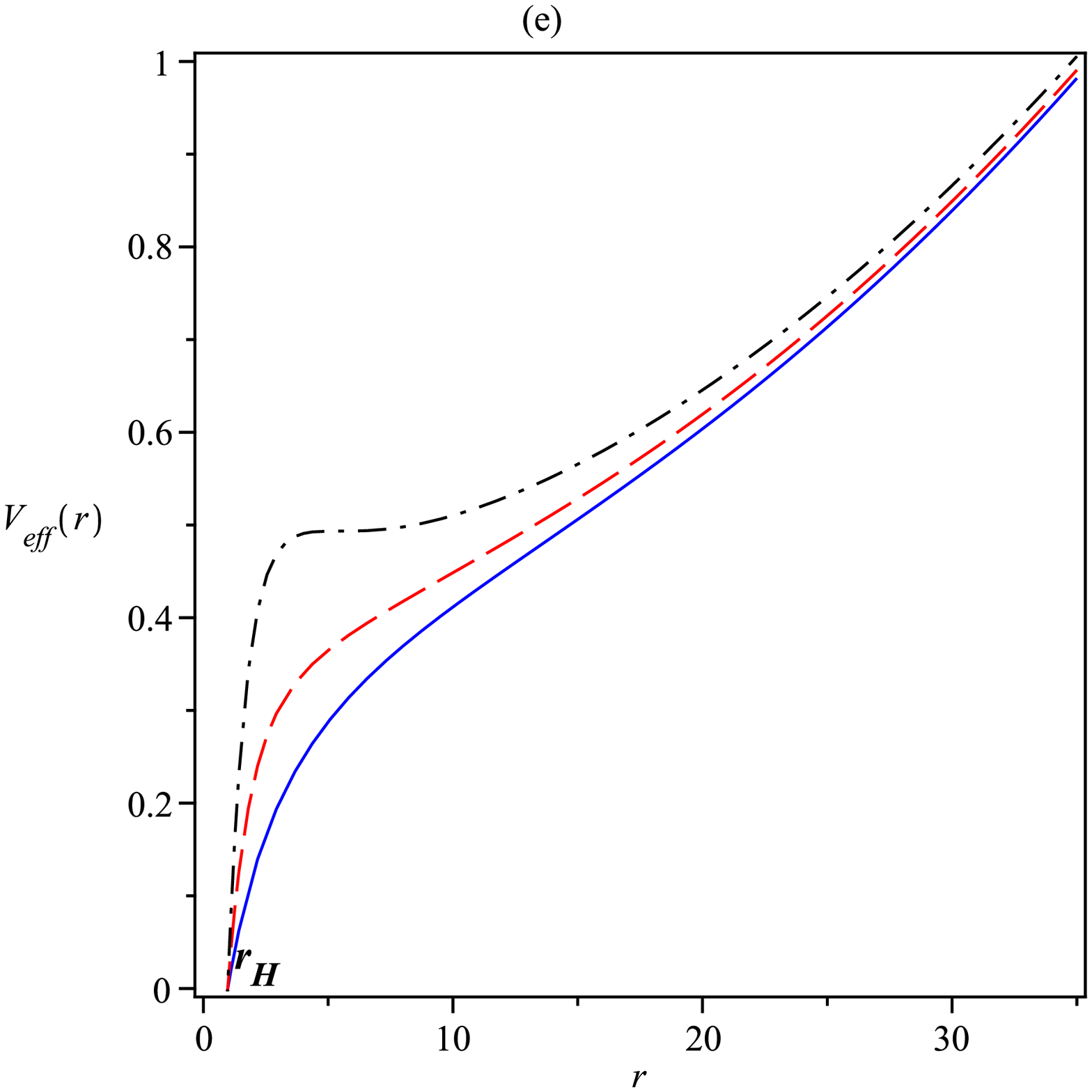}}
\caption{\small{Effective potential for non radial geodesic with $\mu=1$, $g=0.02$, $\beta=1$, $r_{_H}$ represents the position of event horizon. Figs.(a)-(e) represent the cases of $n=0,1,2,3,4$ respectively, where for ssolid line $L=2$, for dashed line $L=4$ and for dot-dashed line $L=6$. \protect\label{fig:potential_nr_02}}}
\end{figure}

\noindent 
Here we study the timelike geodesics for incoming test particles with nonzero angular momentum.
Let us try to understand orbits of the particles in the background of the R-charged black holes by using  the effective potential $V_{eff}$  given by eq.(\ref{eq:Potential_Non_Radial}).
For $r\rightarrow r_{_H}$ (the horizon radius), $V_{eff}=0$ and for $r\rightarrow \infty$, $V_{eff}\rightarrow \frac{1}{2}+\frac{ng^2}{4}\left(\frac{n}{2}-1\right)\mu^2\sinh^2(\beta)+\left(ng^2\mu^2\sinh^2(\beta)\right)\frac{r}{2}+g^2 r^2$, whereas for
$r\rightarrow \infty$, $V_{eff}\rightarrow {1\over{2}}$ (when $g=0$). These are the asymptotic behavior of the potential. In fig.(\ref{fig:potential_nr_01}) and fig.(\ref{fig:potential_nr_02}), the $V_{eff}$ are given in several plots varying the charge parameters $\beta$ and the angular momentum $L$ respectively keeping all the other parameters fixed. 
One can observe from figs.(\ref{fig:potential_nr_01}a)-(\ref{fig:potential_nr_01}e), that for a particular value of $n$ the height of the potential decreases with the increasing values of the charges. One can also notice that height will become small in fig.(\ref{fig:potential_nr_02}) when the values of the angular momentum are small for a particular value of $n$.

\newpage
\begin{figure}[h]
\centerline{\includegraphics[scale=0.25]{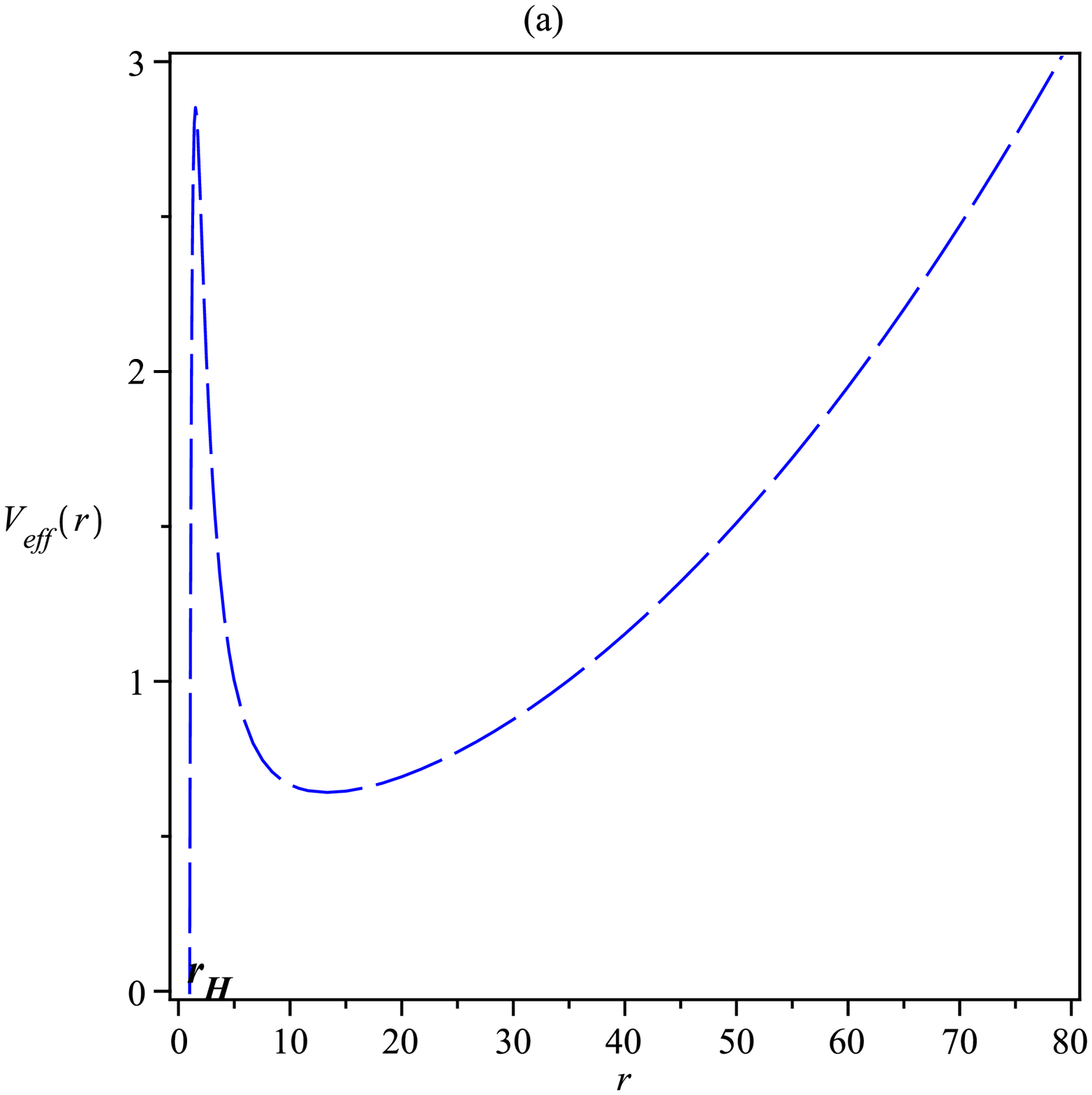}
\hskip-1mm \includegraphics[scale=0.25]{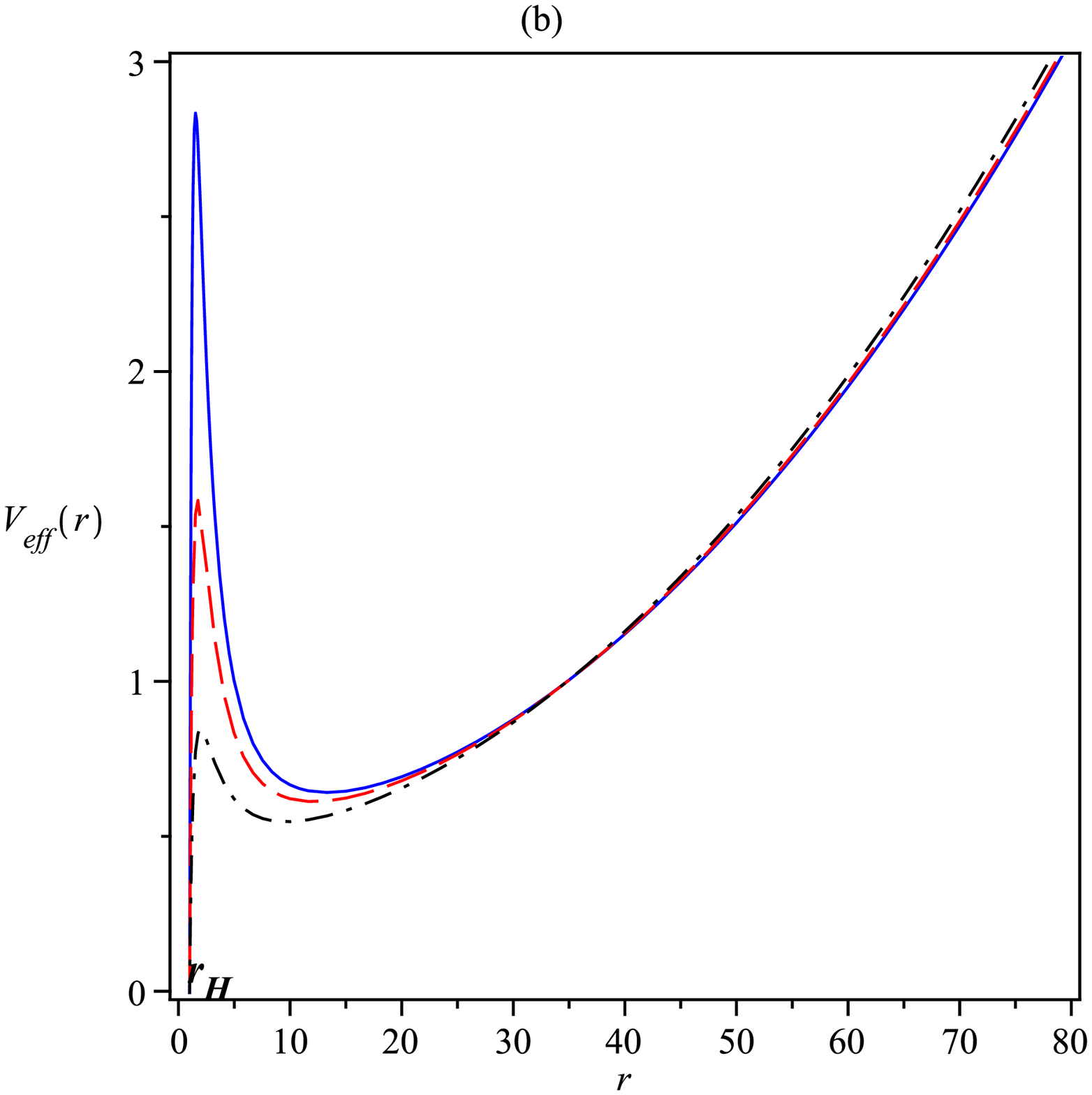}
\hskip-1mm \includegraphics[scale=0.25]{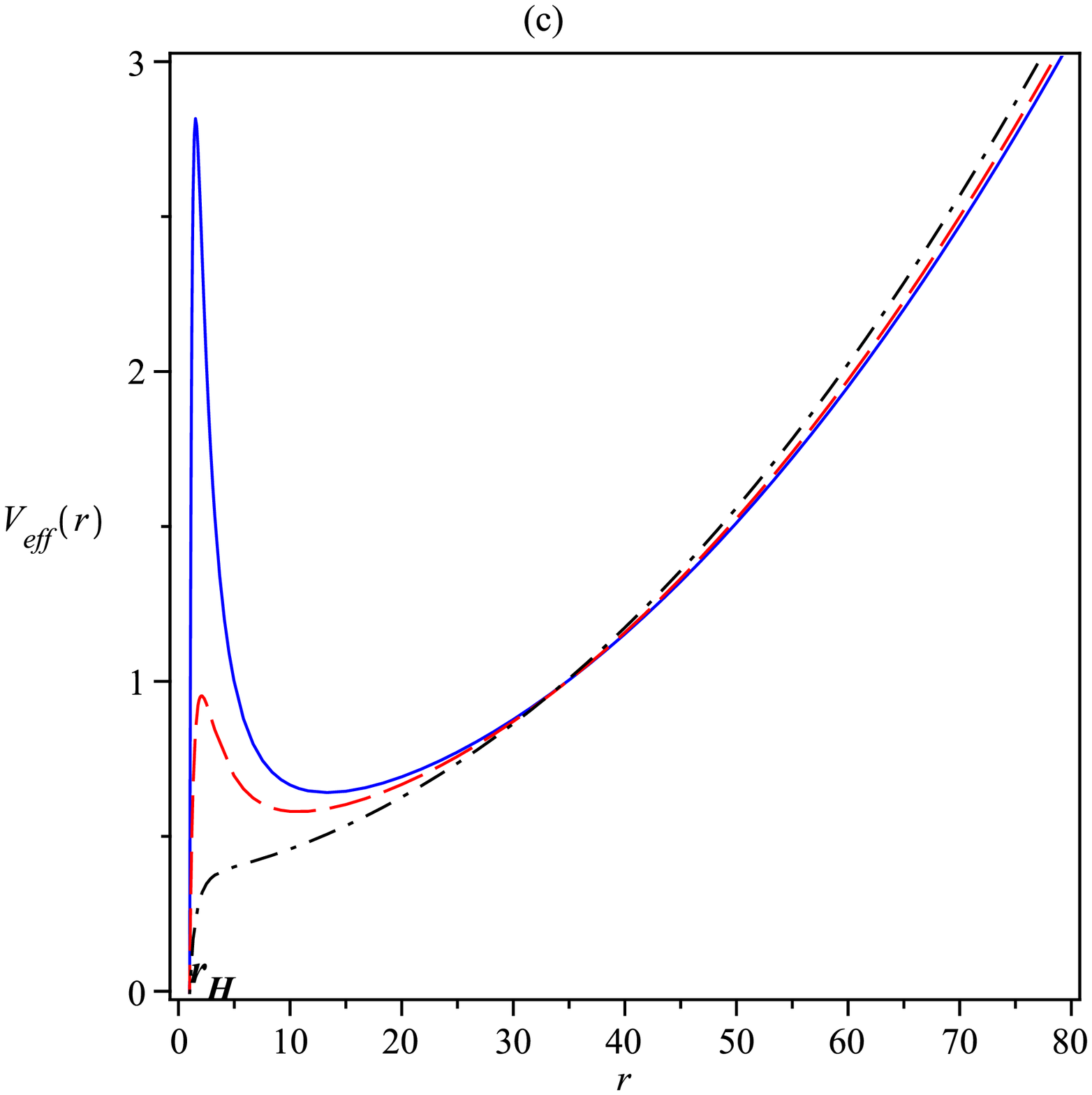}}
\nonumber
\end{figure}
\begin{figure}[h]
\centerline{\includegraphics[scale=0.25]{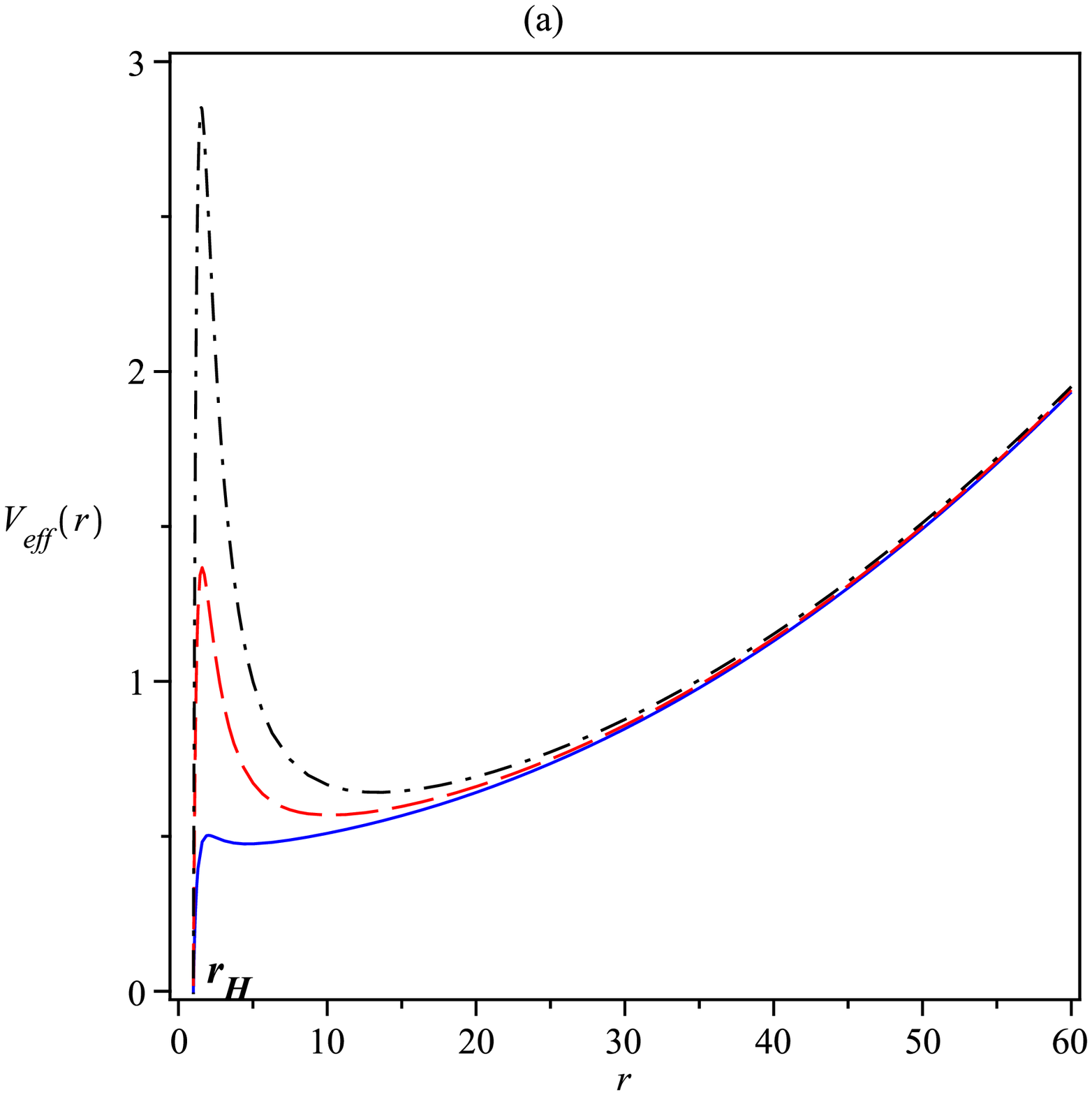}
\hskip-1mm \includegraphics[scale=0.25]{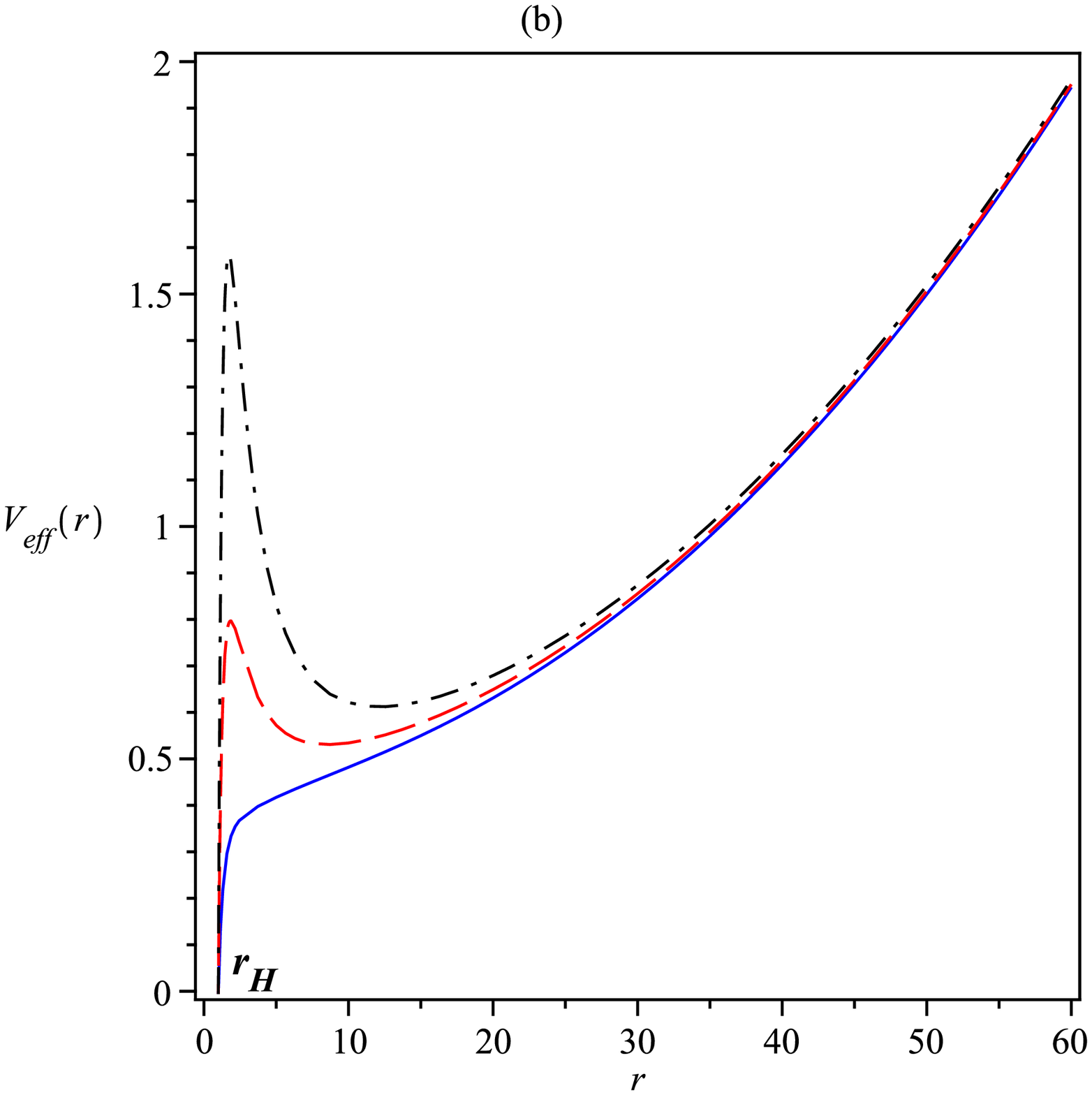}
\hskip-1mm \includegraphics[scale=0.25]{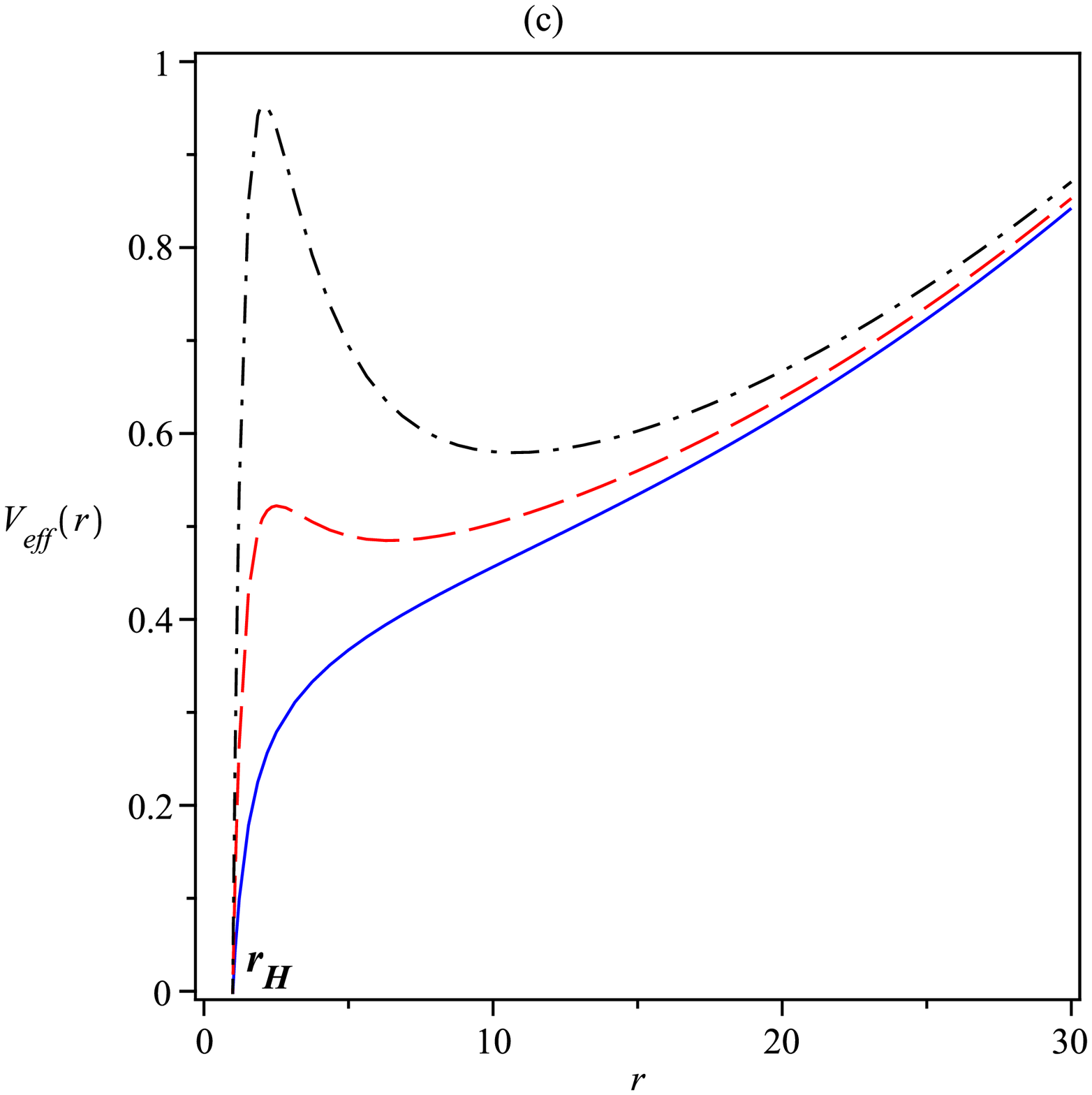}}
\nonumber
\end{figure}
\newpage
\begin{figure}[h]
\centerline{\includegraphics[scale=0.4]{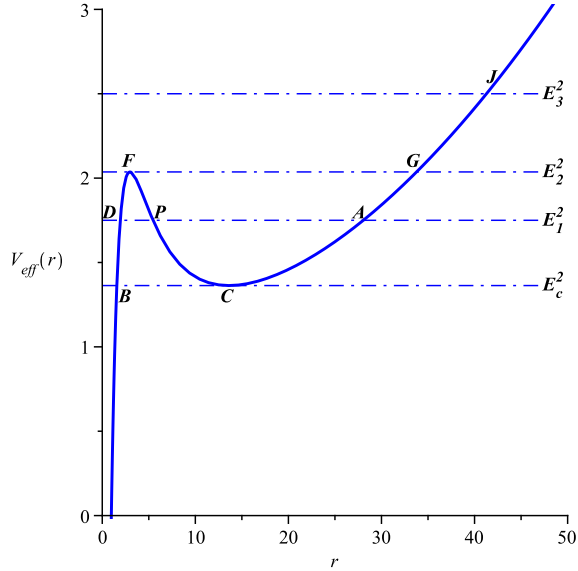}}

\vspace*{4pt}
\caption{\small{Effective potential for unit mass black hole (i.e. $\mu=1$) in the presence of four equal charges with $L=10$, $\beta$=1, $g=0.02$.
\protect\label{potential_4n_gen}}}
\end{figure}
\noindent 
The following orbits are allowed depending on the values of the constant $E$ (i.e. energy of the incoming test particle) as plotted in fig.(\ref{potential_4n_gen}).\\
I) \underline{$E=E_c$:}\\
(i) Here $V_{eff}={E_c}^2$ and $\dot{r}=0$ leading to a {\it stable circular orbit} at point $C$ of fig(\ref{potential_4n_gen}).\\
(ii) For the other possible orbit at this energy value, the test particle starts from the point $B$ and falls into the singularity. Hence it is a {\it terminating bound orbit}.\\
II) \underline{$E=E_1$:}\\
(i) A bounded {\it planetary orbit} between points $A$ and $P$ (as shown in fig(\ref{potential_4n_gen})).\\
(ii) The other possible orbit corresponds to point $D$ which falls into the singularity (i.e. a {\it terminating bound orbit}).\\
III) \underline{$E=E_2$:}\\
The possible orbits are {\it unstable circular orbit} at point $F$. 
The particle starts from point $F$ and then it can go either to point $G$ or to the singularity after crossing the horizon.\\
IV) \underline{$E=E_3$:}\\
There exists {\it terminating bound orbit} for particle crossing point $J$.
Hence there are no fly-by orbits possible.
\subsection{Analysis of the geodesics:}
\begin{figure}[h]
\centerline{\includegraphics[scale=0.3]{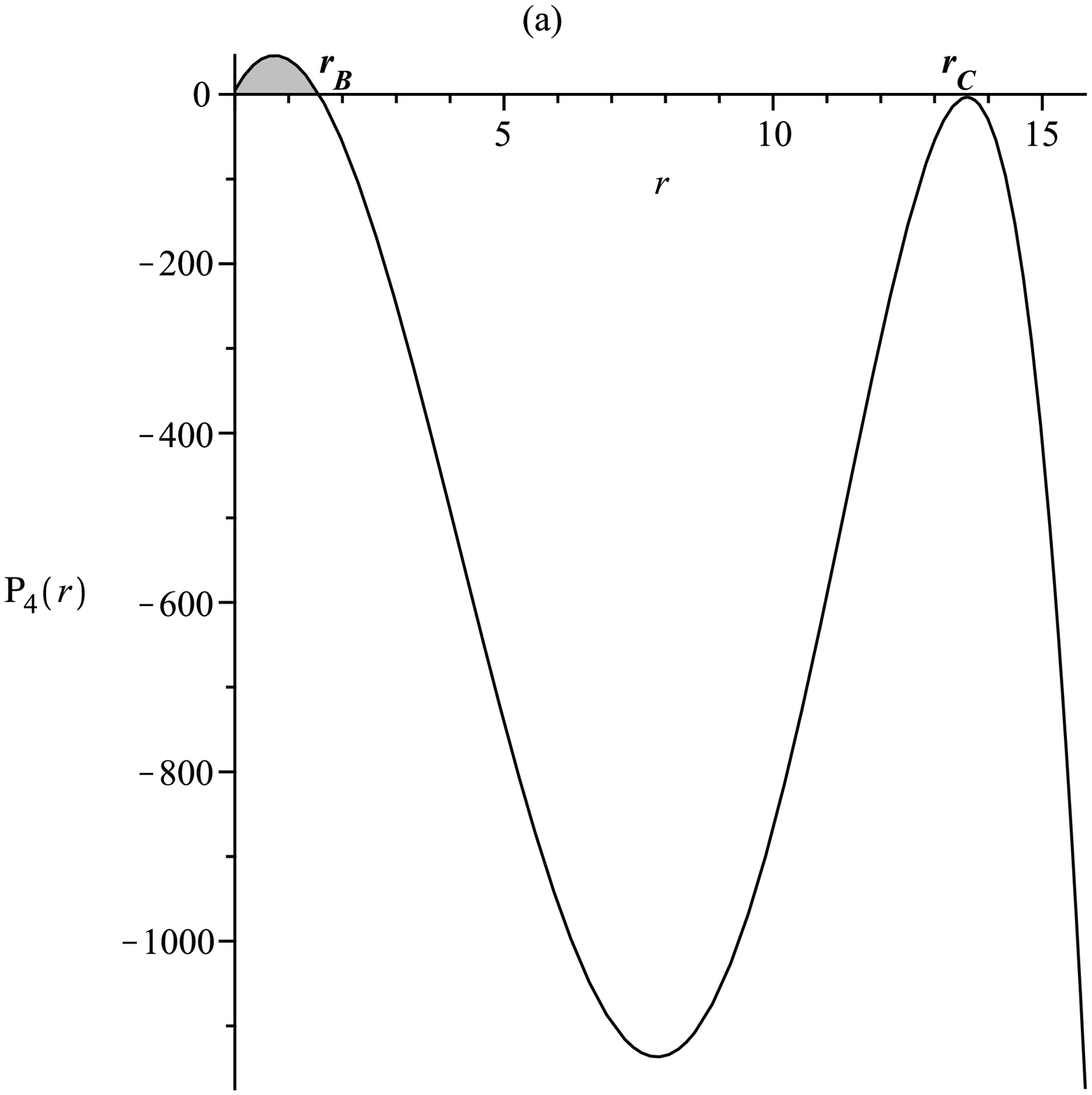}
\hskip6mm \includegraphics[scale=0.3]{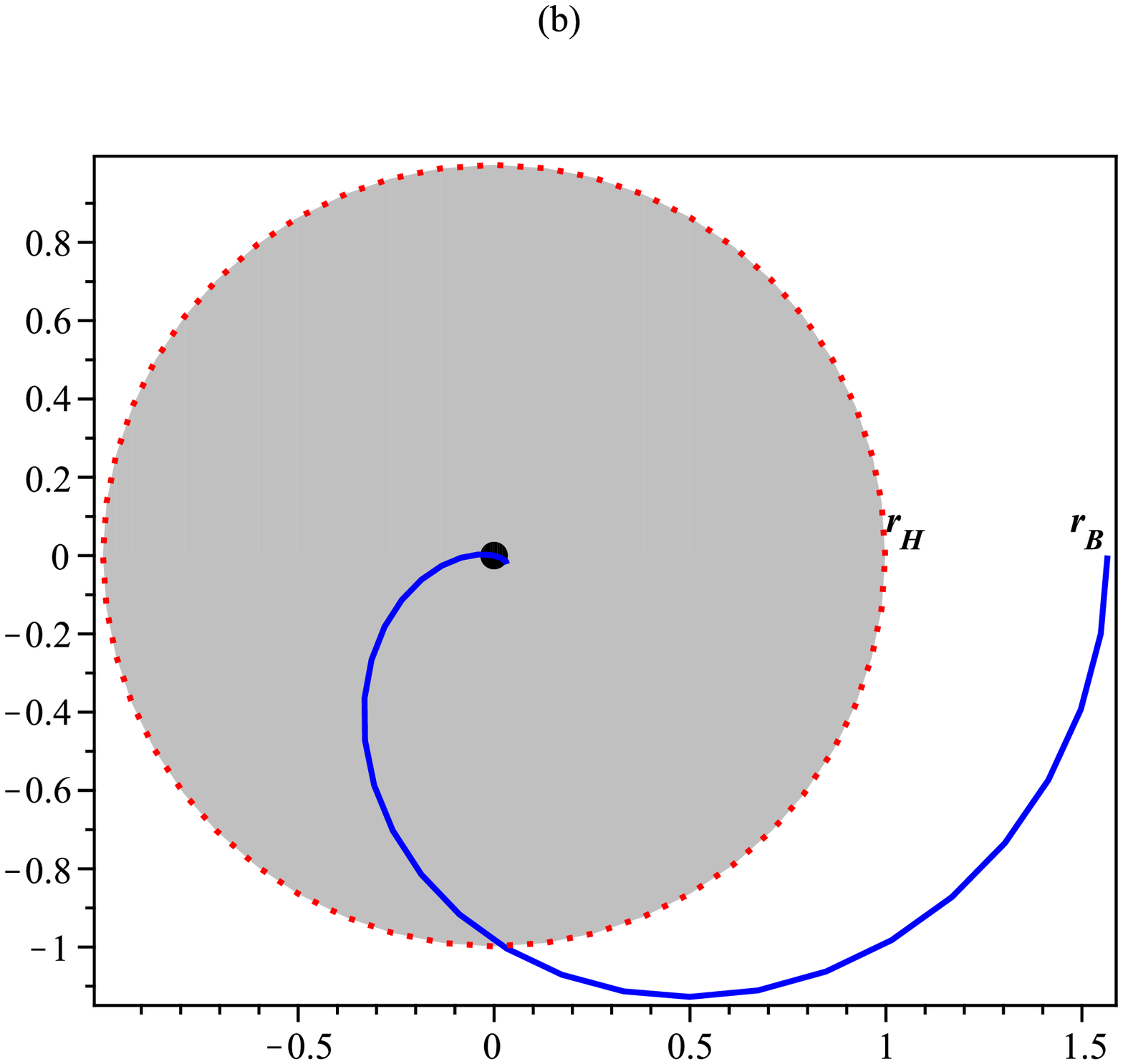}}
\vspace*{8pt}
\caption{\small{{\bf (a)} $P_4(r)$ representing the shaded allowed region for motion of test particle with a positive real zero at point $B$ and a degenerate root at point $C$(where the {\it stable circular orbit} exists), 
{\bf (b)} Solid line represents the orbit ({\it terminating bound orbit}) for test particle starting from a distance at $r=r_B$ in the allowed region, dotted line represents the event horizon at $r=r_H$; with $L=10$, $\beta$=1, $g=0.02$, $E^2=1.3632$.}}
\protect\label{circular}
\end{figure}

\noindent Orbit equation can be obtained by using eq.(\ref{Com2}) and eq.(\ref{Eom}) as,

\begin{equation}
\left(\frac{dr}{d\phi}\right)^2=\frac{1}{L^2}{P_n}(r),
\protect\label{Orbital}
\end{equation}
\begin{equation}
{P_n}(r)=\left(E^2 H^n-f H^{n/2}-\frac{L^2 f}{r^2}\right)r^4.
\end{equation}
where, $f$ and $H$ are given in eq.(\ref{eq:f_n_equal_q}). 
The physically acceptable regions having positive values of $P_n(r)$ or equivalently $E^2\ge{V_{eff}}$ for positive and real values of $r$, represents the allowed region of motion for test particles. Hence the number of positive real zeros of $P_n(r)$ uniquely determine the type of particles orbit in the background of these charged black holes. For $n=4$, the polynomial $P_n(r)$ reduces to,
\begin{equation}
P_4(r)=E^2\left(r+{p}\right)^4-\left[L^2+\left(r+{p}\right)^2\right]\left[r^2-r+2g^2\left(r+{p}\right)^4\right],
\end{equation}
where the mass parameter is fixed as $\mu=1$.
Note that for any value of the parameters $\beta$, $g$, $L$, $E$, the function 
$P_4(r)\rightarrow{-\infty}$ for $r\rightarrow{\pm\infty}$. Also, for $r=0$, 
$P_4(r)= {2g^2 p^4\over{L^2}}({E^2\over{2g^2}}-L^2-p^2)$ indicates that $P_4(r)$ becomes positive in the domain of the parametric space where ${E^2\over{2g^2}}> L^2+p^2$. 
We restrict the values of the parameters in such a way that the previous bound satisfied throughout hereafter.\\
$\bf(I)$ \underline{\bf For E = $E_C$:}\\
As discussed above that for energy value $E = E_C$,  two type of orbits are possible, one is a stable circular orbit and another one is the terminating bound orbit. In fact  the number of zeros of the function $P_4(r)$ uniquely characterizes the orbit structure as mentioned in {\cite{Heckm2008}}. Fig.(\ref {circular}) represents these cases numerically.
   
\begin{figure}[h]
\centerline{\includegraphics[scale=0.3]{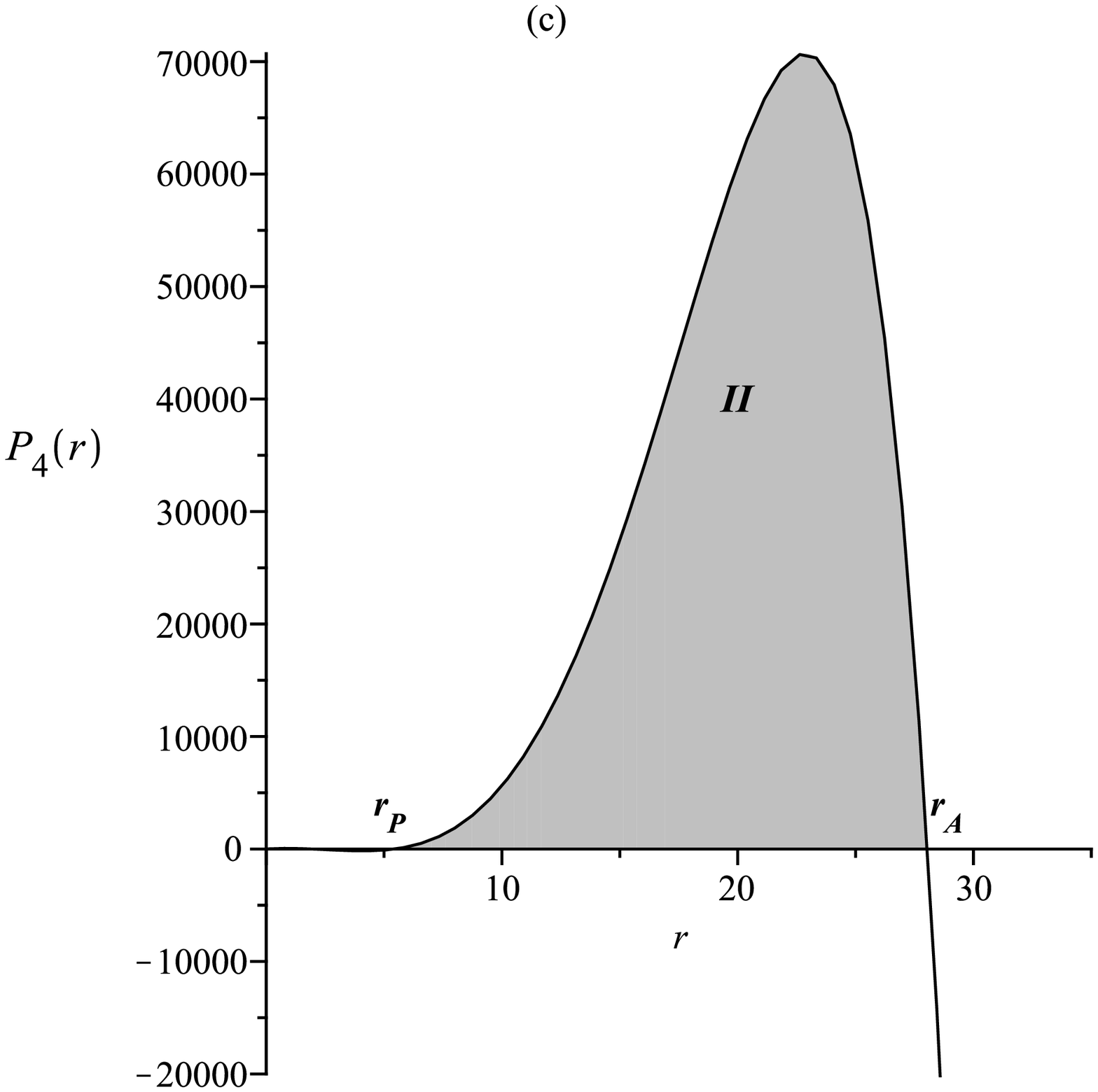}
\hskip5mm \includegraphics[scale=0.3]{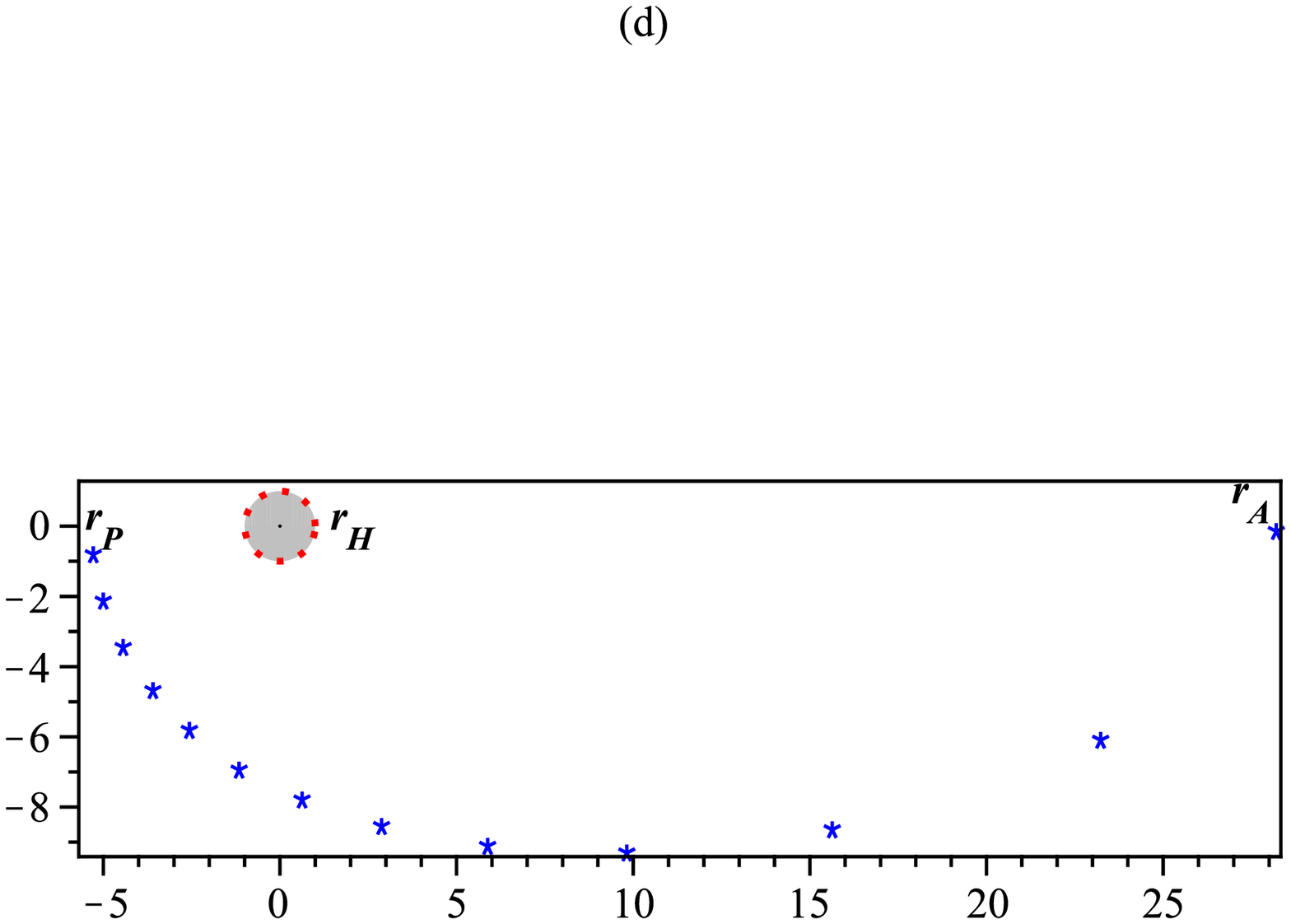}
}
\vspace*{8pt}
\caption{\small{{\bf (a)$\&$(c)} $P_4(r)$ representing the shaded allowed region for motion of test particle with three real positive zeros at points $\bf{D,P,A}$, 
{\bf (b)} Solid line represents the orbit for test particle starting from point $D$ in the allowed region and dotted line represents the event horizon, {\bf (d)} Asterisk curve represents the orbit of test particle bounded between points $A$ and $P$ while dotted line represents the event horizon, 
; with $L=10$, $\beta$=1, $g=0.02$, $E^2=1.75$, $n=4$. ({\bf (a)} \& {\bf (c)} represent two magnified parts of a single plot).}}
\protect\label{E_E_1}
\end{figure}

\begin{figure}[h]
\centerline{\includegraphics[scale=0.3]{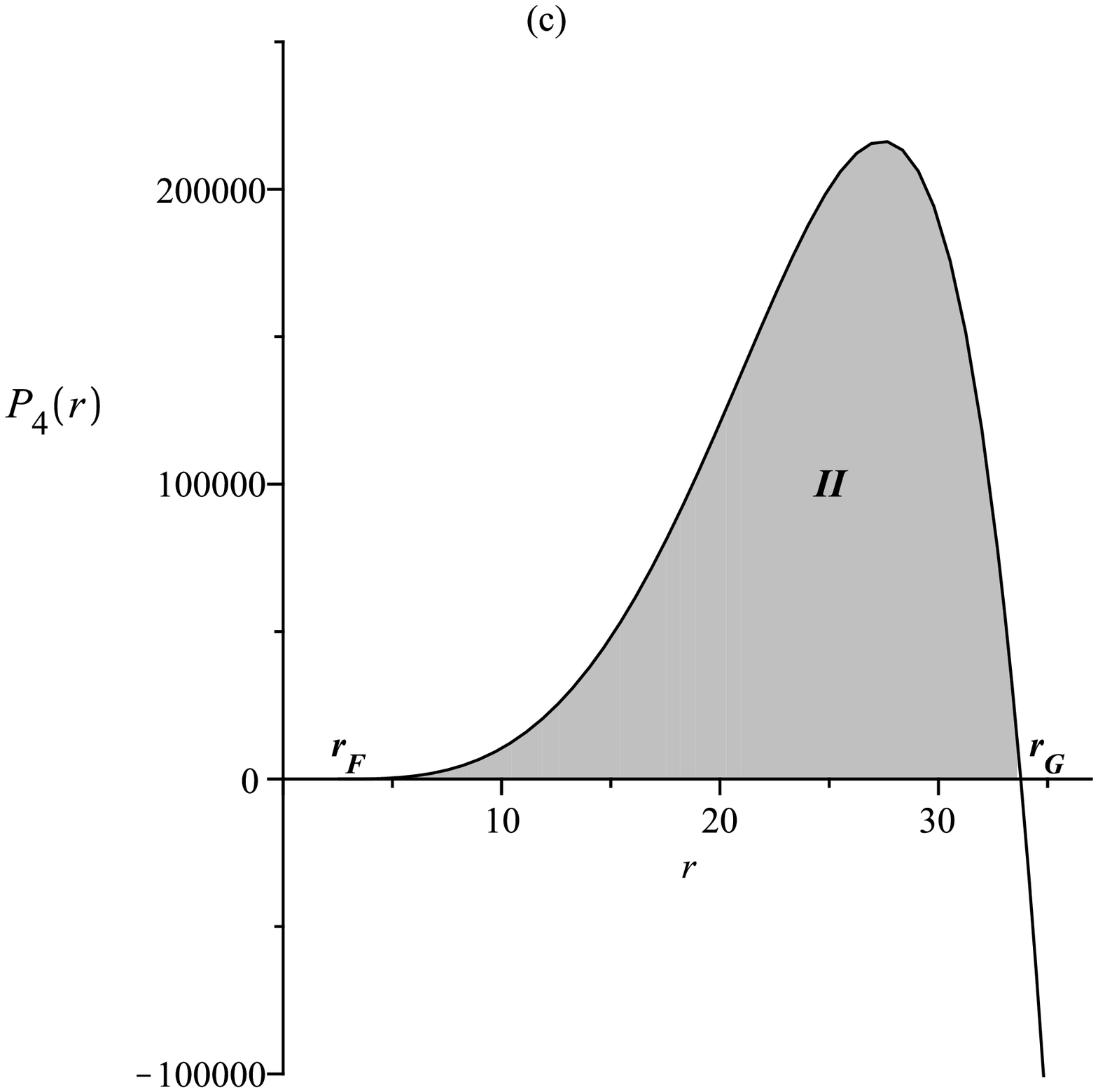}
\hskip5mm \includegraphics[scale=0.3]{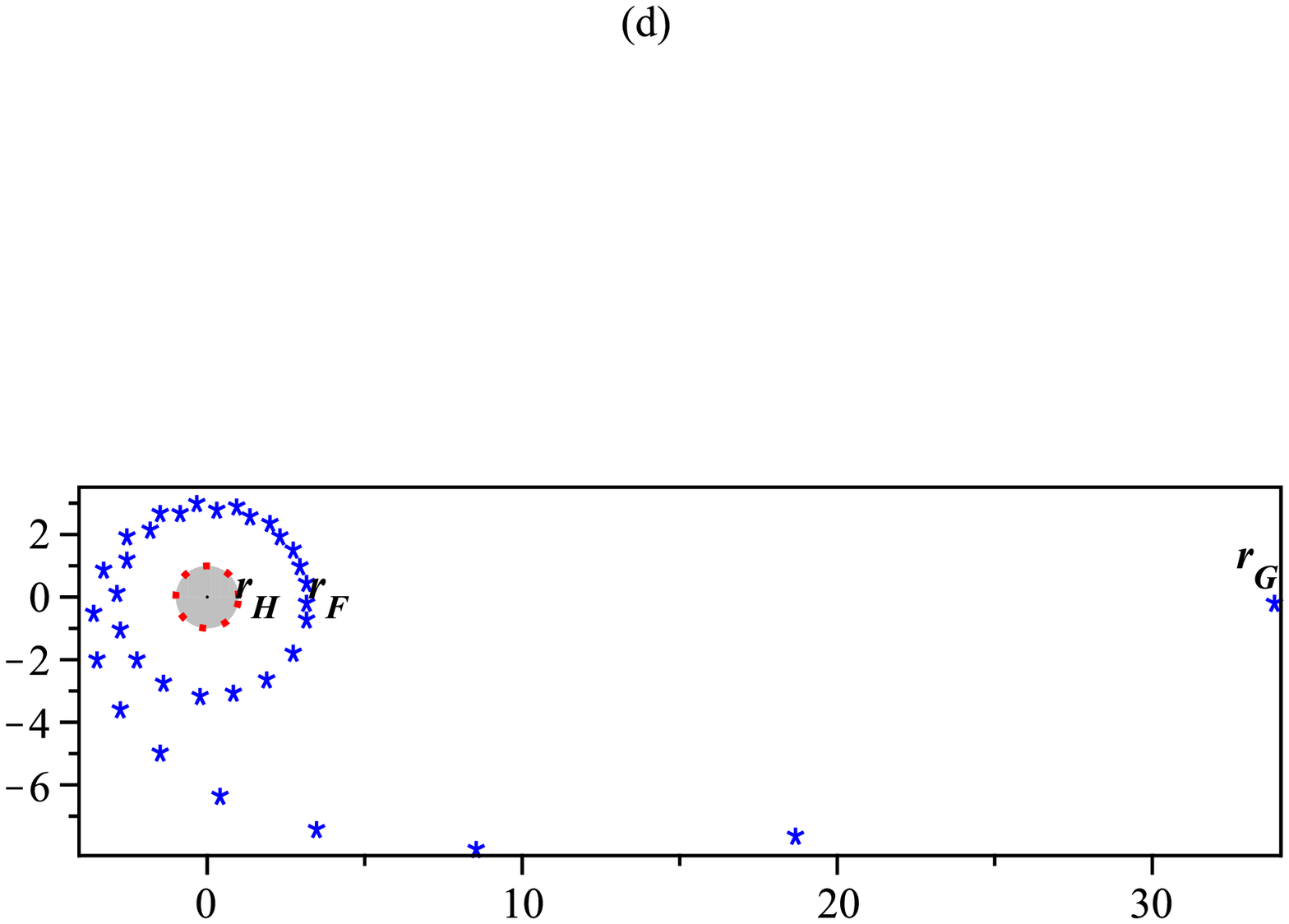}}
\vspace*{8pt}
\caption{\small{{\bf (a)$\&$(c)} $P_4(r)$ representing the shaded allowed region for motion of test particle with two real positive zeros at points $\bf{F}$ and $\bf{G}$, 
{\bf (b)} Solid line represents the {\it terminating bound orbit} for test particle starting from point $F$ to the singularity and dotted line represents the event horizon, {\bf (d)} Asterisk curve represents the {\it unstable circular orbit} of test particle while dotted line represents the event horizon; with $L=10$, $\beta$=1, $g=0.02$, $E^2=2.0366$, $n=4$.  ({\bf (a)} \& {\bf (c)} represent two magnified parts of a single plot).}}
\protect\label{E_E_2}
\end{figure}

\begin{figure}[h]
\centerline {\includegraphics[scale=0.3]{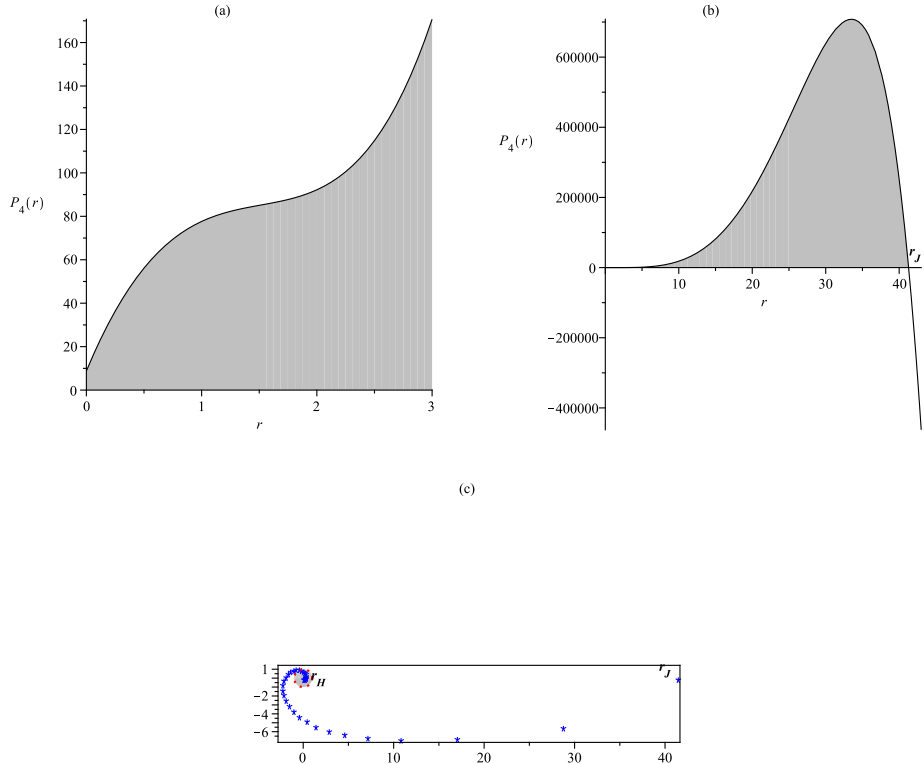}}
\vspace*{4pt}
\caption{\small{{\bf (a) \& \bf (b)} $P_4(r)$ representing the shaded allowed region for motion of test particle with one real positive zeros at point $\bf{J}$,
{\bf (c)} Asterisk curve represents the {\it terminating bound orbit} of test particle while dotted line represents the event horizon; with $L=10$, $\beta$=1, $g=0.02$, $E^2=2.5$, $n=4$. ({\bf (a)} \& {\bf (b)} represent two magnified parts of a single plot).}}
\protect\label{E_E_3}
\end{figure}

\noindent
\underline{\it Circular Orbits:}\\
In case of {\it circular orbit}, $r=r_C=$ constant, where $r_C$ is the distance of the circular orbit from the singularity and hence $\dot{r}=0$.
In order to calculate the time periods for circular orbits by using eqs.(\ref{Com1}) and (\ref{Com2}), we have
\begin{equation}
dt = \frac{E\mathcal{H}(r_C)}{Lf(r_C)}{r^2_C} d\phi.
\label{eq:t-phi}
\end{equation}
The circular orbit condition ${V^{\prime}_{eff}}=0$ can be used to express $L$ in terms of $r_C$ as,
\begin{equation}
L^2=\frac{\mathcal{X}}{\mathcal{Y}},
\end{equation}
where $\mathcal{X}=\frac{\mu np}{2r^3_C}-\frac{1}{r^2_C}\left[\frac{np}{2}+\mu\left(1+\frac{p}{r_C}\right)\right]-2g^2\mathcal{H}(r_C)\left(-\frac{np}{2}+2p+2r_C\right)$,\\
$\mathcal{Y}=\mathcal{H}^{-1}(r_C)\left(-\frac{2}{r^3_C}+\frac{3\mu}{r^4_C}+\frac{np}{r^4_c{H}(r_C)}-\frac{\mu np}{r^5_c{H}(r_C)}\right)+\frac{4g^2}{r_C{H}(r_C)}\left(1-{H}(r_C)+\frac{p}{r_C}\right)$,
\\$p=\mu\sinh^2(\beta)$, $H=1+\frac{p}{r_C}$ and $\mathcal{H}=H^n$. \\
As radial velocity $\dot{r}$ vanishes for circular orbit, it provides another condition for this orbit as $E^2=V_{eff}(r_C)$. Substituting this in eq.(\ref{eq:t-phi}) with $\Delta t\equiv{T}_t$ and $\Delta\phi=2\pi$ for one period then we obtain,
\begin{equation}
T_t=\left[\frac{\mu r_C}{2f(r_C)}\left(\frac{1}{r^2_C}+\frac{\mathcal{H}^{1/2}(r_C)}{L^2(r_C)}\right)\right]^{1/2}\left(\frac{r_C}{r_C+p}\right)T_{t,sch},
\label{q:Tt-n}
\end{equation}
where $T_{t,sch}=2\pi\sqrt{\frac{2r^3_C}{\mu}}$. 
Similarly time period in proper time can also be obtained using eqs.(\ref{Com2}) and (\ref{eq:t-phi}) as,
\begin{equation}
T_{\tau}=\left[\frac{\mu r_C}{2f(r_C)}\left(\frac{1}{r^2_C}+\frac{\mathcal{H}^{1/2}(r_C)}{L^2(r_C)}\right)\left(\frac{2r_C}{\mu}-3\right)\right]^{1/2}\left(\frac{r_C}{r_C+p}\right)2\pi{r^2_C}
\nonumber
\end{equation}
\begin{equation}
={r_C}\left[\frac{\mu r_C}{2f(r_C)}\left(\frac{1}{r^2_C}+\frac{\mathcal{H}^{1/2}(r_C)}{L^2(r_C)}\right)\right]^{1/2}T_{\tau,sch}.
\label{q:T_tau}
\end{equation}
Now we comment on the values computed for time periods $T_t$ and $T_\tau$ respectively from eqs. (\ref{q:Tt-n}) and (\ref{q:T_tau}). 
One can also easily calculate the numerical values of time periods for a definite set of parameter and  a comparison of  $T_t$ and $T_{\tau}$ is presented in table {\ref{table:1}}.
\begin{table}[h!]
\centering
\begin{tabular}{ | c | c | c | c | }
\hline
$n$ & $r_C$ & $T_t$ & $T_\tau$ 
\\
\hline
\hline
0 & 3.7871 & 2.4851 & $41.6\pi$ \\
\hline
1 & 4.3042 & 1.9965 & $44.42\pi$  \\
\hline
2 & 4.8169 & 1.4322 &  $51.20\pi$  \\
\hline
3 & 5.3145 & 1.0887 &   $58.26\pi$ \\
\hline
4 & 5.7911 & 0.8640 &    $65.43\pi$   \\
\hline
\end{tabular} 
\caption{Comparative view of the time periods for different number of charges for $\mu=1$, 
$g=0.02$ and $\beta=1$.}
\label{table:1}
\end{table}
In comparison, one can see the effect of charges on the time periods and also on the radius of circular orbits, which significantly differ from the corresponding values of the Schwarzschild black hole {\footnote {for Schwarzschild black holes $r_C=3$, $T_t=2.5375$ and $T_\tau=31.18\pi$, for $\mu=1$.}} in GR.
\\
$\bf (II)$ \underline{\bf For E = $E_1$:}\\
In this case, $P_4(r)$ has three positive real zeros. The region $II$ in the fig.(\ref{E_E_1}c) bounded between two roots of $P_4(r)$ at $r_A$ and $r_P$ represents the bound orbit as has been discussed earlier from the effective potential. Another region $I$ in the fig.(\ref{E_E_1}a) gives us the bound orbit where particle should start form the point $r_D$, shown in fig.(\ref {E_E_1}b) and crosses the horizon at $r_H$ and then falls in to the singularity. The trajectories of the particles have been simulated in figs.(\ref{E_E_1}$b$) and (\ref{E_E_1}$d$).
\begin{figure}[h]
\centerline{\includegraphics[scale=0.3]{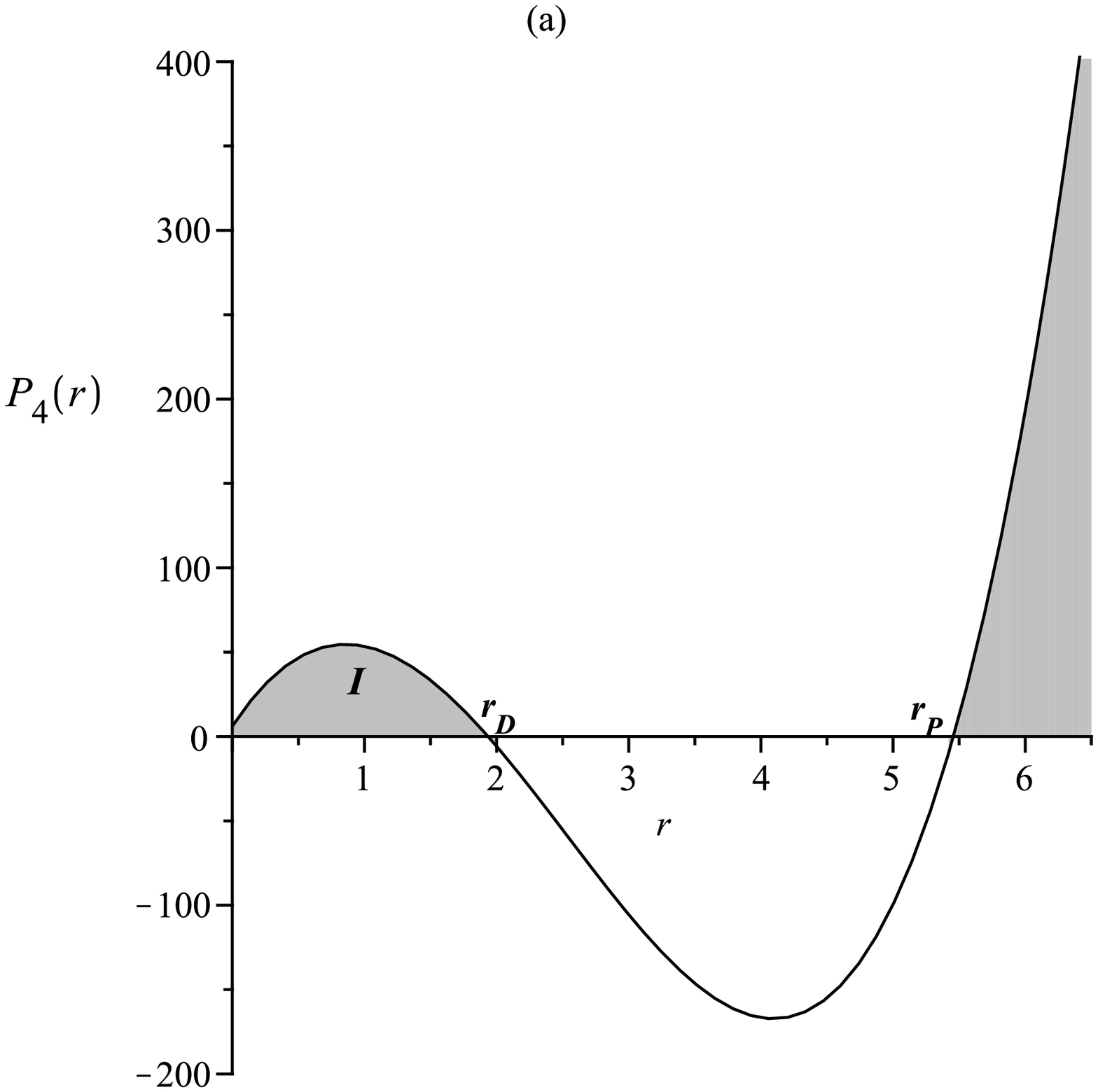}
\hskip5mm \includegraphics[scale=0.3]{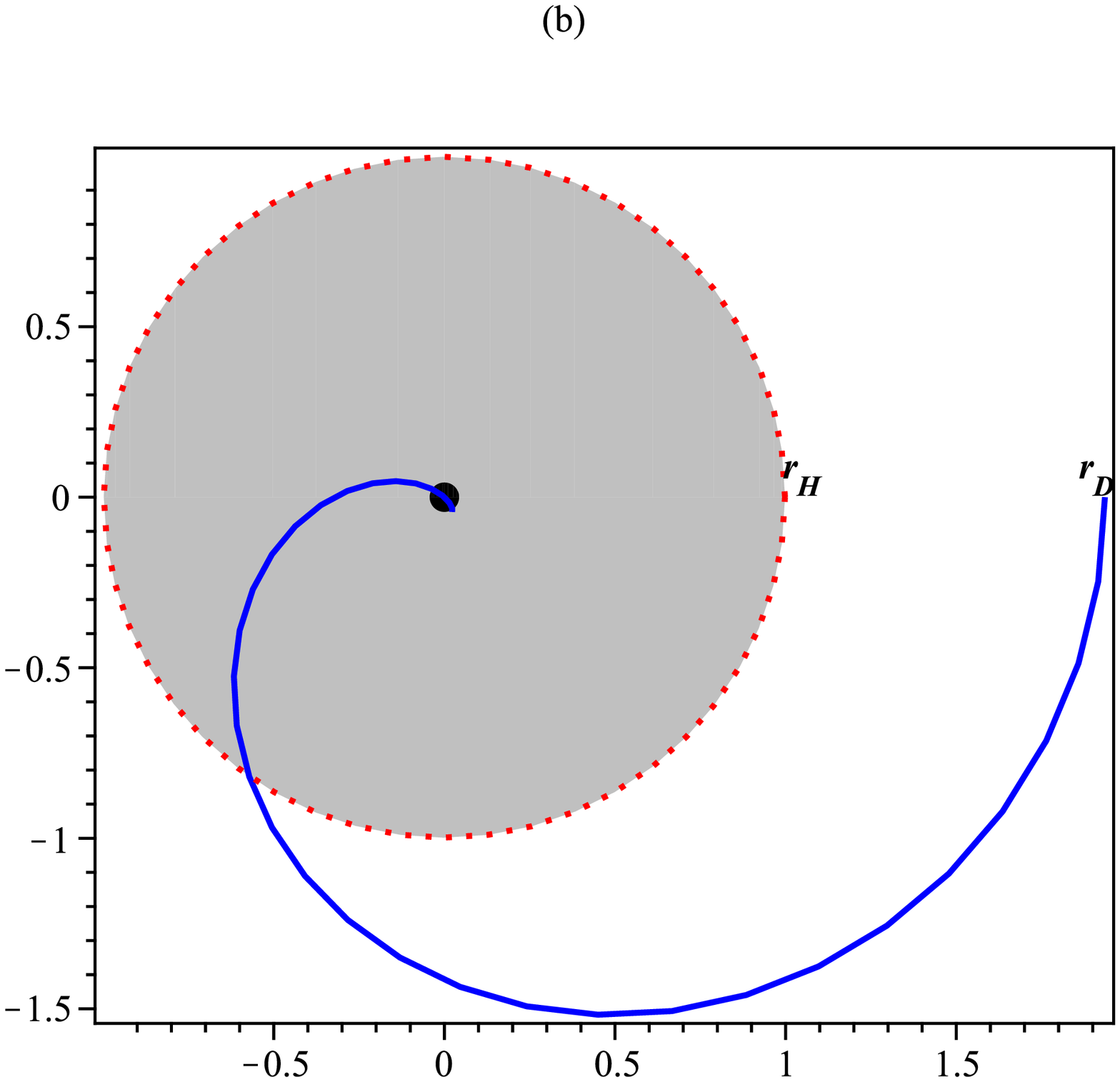}
}
\nonumber
\end{figure}
\newpage\noindent \underline{\it Advance of perihelion for planetary orbits:}\\
We use the elementary derivation of the advance of perihelion of a planetary orbit for the Schwarzschild solution as in \citep{Corn1993}. 
The advance of the perihelion for our case is obtained by comparing a Keplerian ellipse in a
Lorentzian coordinates with the spacetime of interest given in eq.(\ref{met1}). 
The relevant relation communicating the two ellipse is the areal constant of Kepler’s second law.\\
\noindent The following transformation of the coordinates, $r$ and $t$, in the binomial approximation can be obtained for the spacetime used in present study, given by eq.(\ref{met1}),
\begin{equation}
dt^{\prime}=\left[1-\frac{(2\mu+np)}{4r}+g^2{r}\left(1+\frac{np}{r}\right)\left(r-\frac{np}{4}\right)\right]dt,
\label{eq:t-prime}
\end{equation}
\begin{equation}
dr^{\prime}=\left[1+\frac{(2\mu+np)}{4r}-g^2{r}\left(1+\frac{np}{r}\right)\left(r-\frac{np}{4}\right)\right]dr.
\label{eq:r-prime}
\end{equation}
Two elliptical orbits are considered, one the classical Kepler orbit in $r$, $t$ space and other for given spacetime in an $r^{\prime}$, $t^{\prime}$ space. 
In the Lorentz space one has,
\begin{equation}
dA=\int^\rho_0{rdrd\phi},
\end{equation}
and hence the Kepler's second law
\begin{equation}
\frac{dA}{dt}=\frac{1}{2}{\rho^2}\frac{d\phi}{dt}.
\end{equation}
Similarly in given spacetime one has,
\begin{equation}
dA^{\prime}=\int^\rho_0{\mathcal{R}dr^{\prime}d\phi},
\label{dA_prime}
\end{equation}
where $r^{\prime}$ is given in eq.(\ref{eq:r-prime}) and the radial function $\mathcal{R}= r(1+\frac{p}{r})^{\frac{n}{4}}$. The binomial approximation of the radial function $\mathcal{R}$ is
\begin{equation}
\mathcal{R}\approx{\Big(1+\frac{np}{4r}\Big)}.
\label{R_approx}
\end{equation} 
Therefore, applied wherever necessary the binomial approximation and the transformation of coordinate, we obtain the following equation after performing the integration (\ref{dA_prime}) by using eqs. (\ref{eq:r-prime}) and (\ref{R_approx})
\begin{eqnarray}
\frac{dA^{\prime}}{dt^{\prime}} & = & \frac{\rho^2}{2}\Big[\Big(1+\frac{npg^2}{16}(21np+20\mu)-\frac{3}{64}n^4p^4g^4\Big) \nonumber \\ & + &
\frac{1}{\rho}\Big(\frac{3np+6\mu}{4}-\frac{n^2p^2g^2}{64}(9np+2\mu)\Big)
\nonumber \\ & + &
\rho{g^2}\Big(\frac{10np+5\mu}{4}-\frac{35n^3p^3g^2}{64}\Big)
+...\Big]\frac{d\phi}{dt}.
\end{eqnarray} 
For a single orbit,
\begin{eqnarray}
\Delta\phi^{\prime}&=&\int^{\Delta\phi=2\pi}_0\Big[\Big(1+\frac{npg^2}{16}(21np+20\mu)-\frac{3}{64}n^4p^4g^4\Big) \nonumber \\ & + &
\frac{1}{\rho}\Big(\frac{3np+6\mu}{4}-\frac{n^2p^2g^2}{64}(9np+2\mu)\Big)
\nonumber \\ & + &
\rho{g^2}\Big(\frac{10np+5\mu}{4}-\frac{35n^3p^3g^2}{64}\Big)
+...\Big]d\phi.
\label{perihelion}
\end{eqnarray}
The polar form of an ellipse is given by
\begin{equation} 
\rho=\frac{l}{1+e\cos\phi},
\label{ellipse} 
\end{equation}
where $e$ is the eccentricity and $l$ is the {\it semi-latus rectum}. 
Therefore applying the binomial approximation alongwith the integration (\ref{perihelion}) and using the eq. (\ref{ellipse}), we obtain the perihelion shift as
\begin{equation}
\Delta\phi^{\prime}\approx 2\pi+\frac{6\pi\mu}{l}+ \frac{3\pi np}{2l}-
\frac{5\pi\mu}{2}\Big(np+l\Big)g^2-\frac{\pi n^2p^2}{8}
\Big(21+\frac{\mu}{2l}\Big)g^2+......
\label{P_Shift}
\end{equation}
The results we have derived for the perihelion shift of the massive test particle in the space- time of a R-charged black holes where all $n$ (where $n=0,1,...,4$) charges are equal. One can see that the corrections have appeared in eq.(\ref{P_Shift}) due to the charges and the gauge coupling constant $g$ apart form the term depends on mass parameter $\mu$. It is also straight forward to check that the form of the perihelion shift equation(\ref{P_Shift}) nicely goes back to the results obtained for Schwarzschild AdS (i.e. $n=0$, here) presented in {\citep{Cruz2005}}{\footnote {The next order terms are discarded in $l$ as given in equation (\ref{P_Shift}).}}.\\
\vspace{2mm}
$\bf(III)$ \underline{\bf For E = $E_2$:}\\
At this energy value of the incoming particles, the two zeros of the polynomial $P_4(r)$ in the previous plots (fig.(\ref{E_E_1})) merge where $P_4(r)=0$ and also ${d\over dr}P_4(r)=0$ at $r=r_F$. The representative plots have been shown in figs.(\ref{E_E_2}a) and (\ref{E_E_2}c). The radius $r_F$ is the radius of the circular orbit which is unstable. In fig.(\ref{E_E_2}b), a polar plot of the timelike geodesic is given for particle arriving from a finite radius $r_F$ and plunges in to the black hole singularity. Whereas in fig.(\ref{E_E_2}d) another possible orbit has shown where the particles coming from $r_G$ and having an unstable circular orbit at $r=r_F$.  \\    
\begin{figure}[h]
\centerline{\includegraphics[scale=0.3]{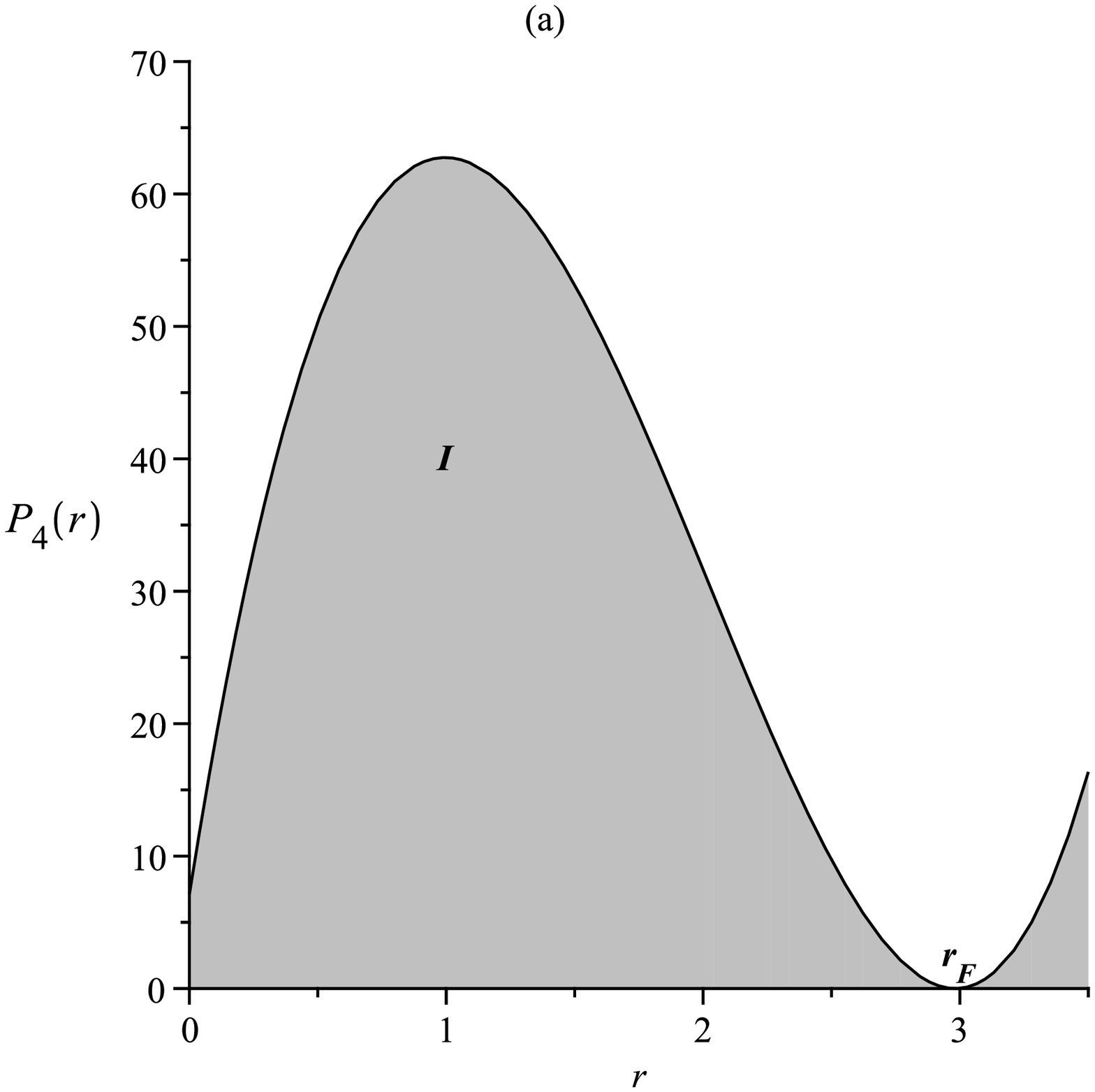}
\hskip5mm \includegraphics[scale=0.3]{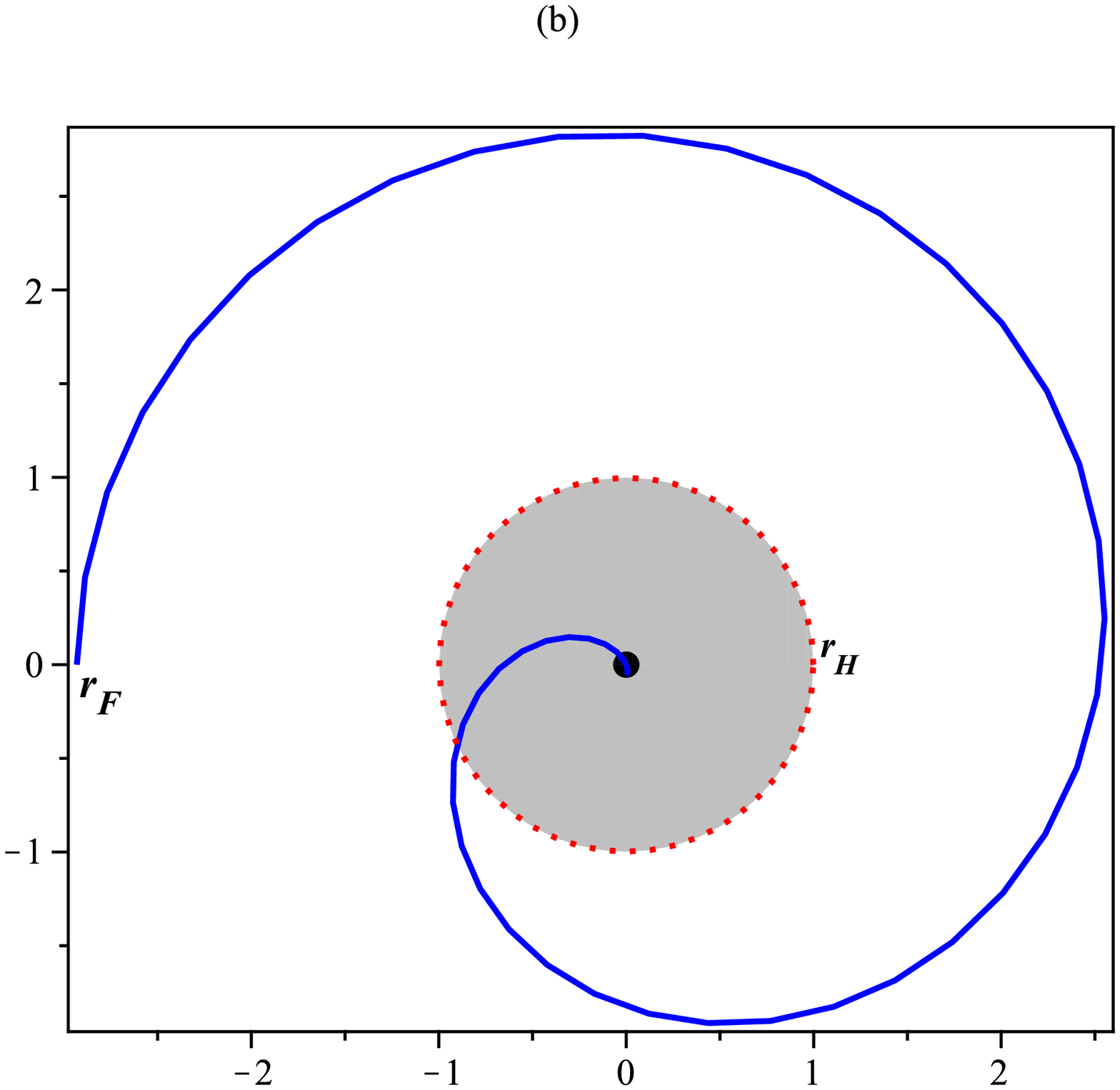}}
\nonumber
\end{figure}
\\
\newpage
\noindent
$\bf(IV)$ \underline{\bf For E = $E_3$:}\\
At this energy value $E=E_3$, the degenerate root at $r=r_F$ will disappear leaving only a single root at $r=r_J$ as shown in fig.(\ref{E_E_3}$b$). Again we have numerically simulated the path of the particles carrying energy $E=E_3$ is shown in fig.(\ref{E_E_3}$c$) represents a terminating bound orbit.\\
\begin{figure}[h]
\centerline{\includegraphics[scale=0.3]{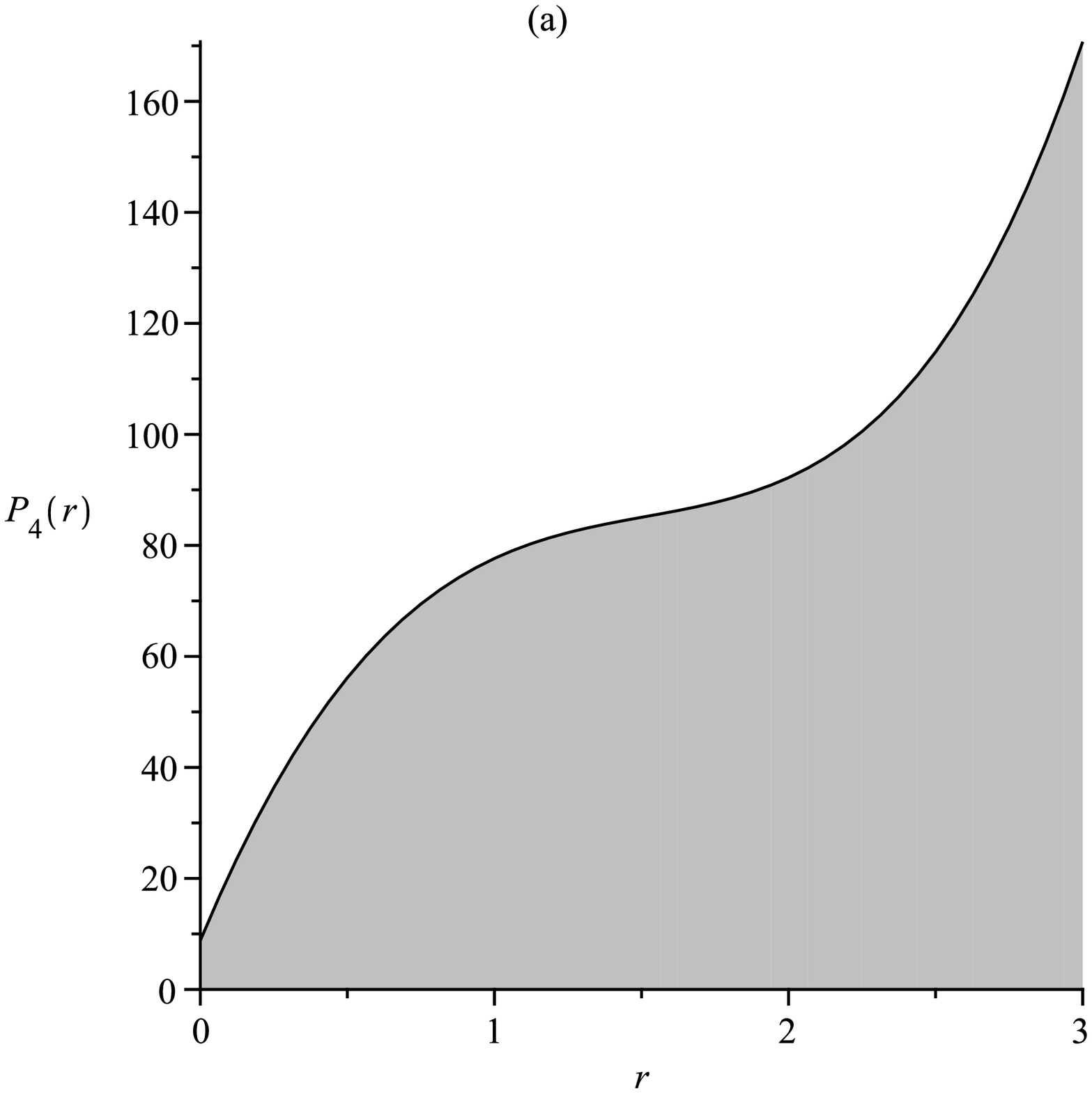}
\hskip5mm \includegraphics[scale=0.3]{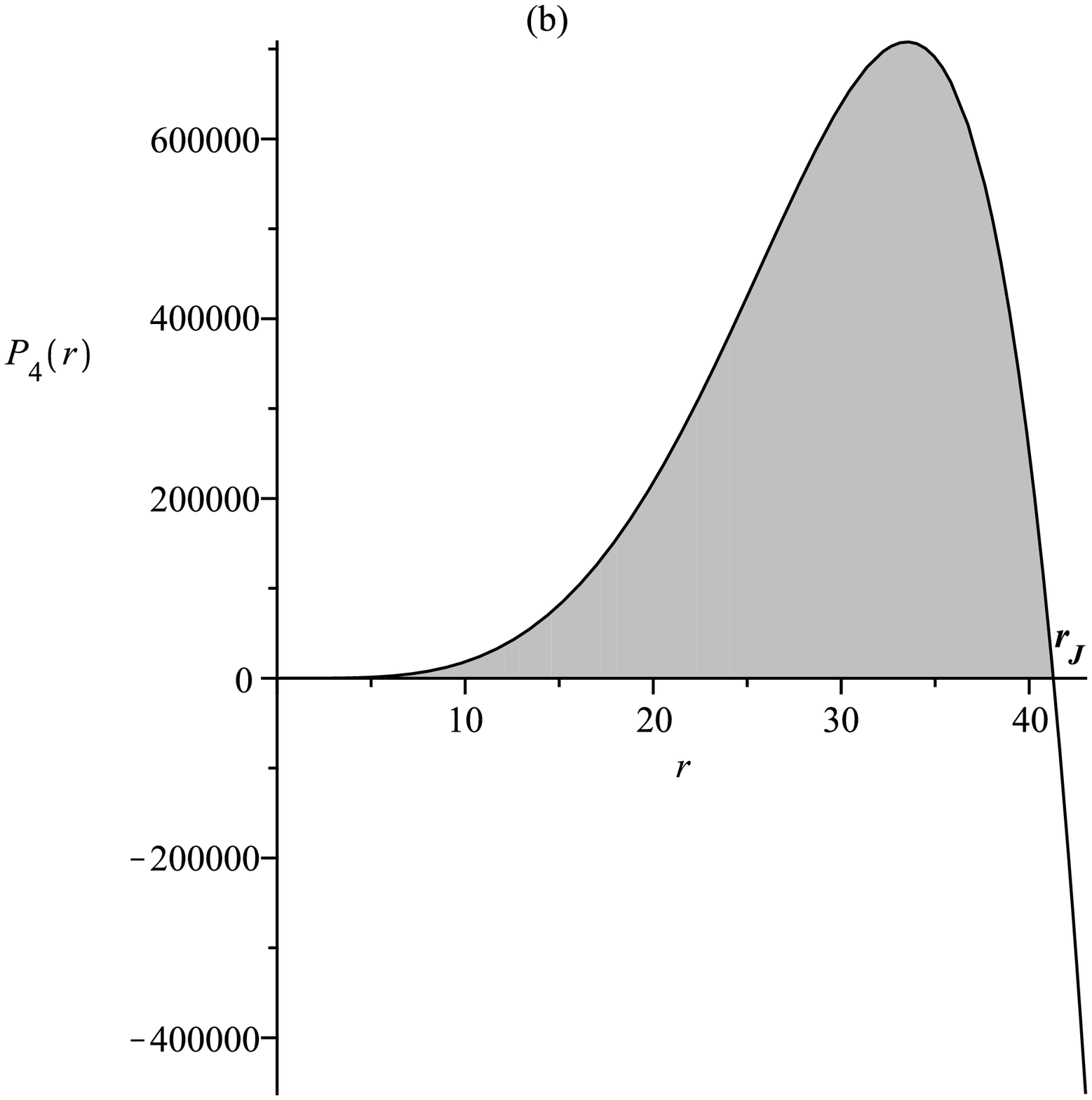}}
\nonumber
\end{figure}
\\
\newpage
\noindent In order to have a detailed view of the geodesic motion in the background of R-charged Black Holes, the Null geodesics are discussed in the next section.
\section{Nature of effective potential and classification of orbits for Null Geodesics}
\subsection{Radial Geodesics}
\begin{figure}[h!]
\centerline {\includegraphics[scale=0.36]{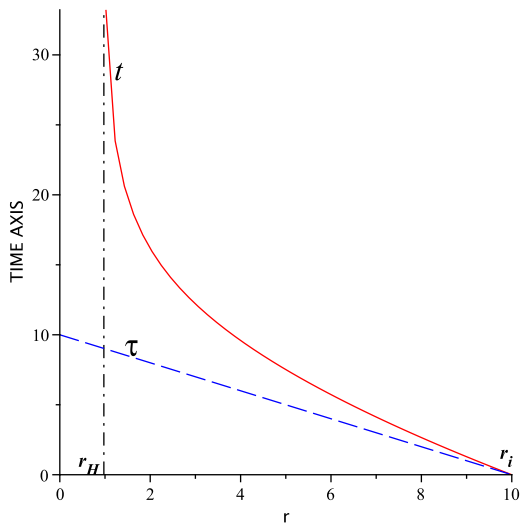}}
\vspace*{4pt}
\caption{\small{Radial motion of an incoming massless test particle for proper time, $\tau$ (dashed curve) and the coordinate time, $t$ (solid curve) with $\mu$ = 1, $E$ = 1, $\beta$ = 1, $n$ = 4, $g$ = 0.02; where $r_H$ ( verticle dot-dashed line) and $r_i$ represent the position of event horizon and the initial position of test particle respectively.}}
\protect\label{time-r-null}
\end{figure}

\noindent Since the radial geodesics are the trajectories followed by zero angular momentum test particles(i.e. L=0) in given spacetime geometry, the effective potential given by eq.(\ref{eq:Potential_Non_Radial_Null}) for radial null geodesics vanishes i.e. $V_{eff}=0$ and consequently the radial equation of motion reduces to,
\begin{equation}
\frac{dr}{d \tau}=\pm E,
\label{r-tau-radial}
\end{equation}
where positive sign corresponds to the outgoing test particles.
The integration of eq.(\ref{r-tau-radial}) leads to,
\begin{equation}
r=\pm{E}\tau+\tau_0,
\label{r-tau-radial-solution}
\end{equation}
where $\tau_0$ is the integration constant corresponding to the initial position of the test particle.
Eq.(\ref{r-tau-radial-solution}) shows that in terms of proper time, the radial coordinate depends only on the constant energy value $E$.\\
\noindent Using eq.(\ref{Com1}) and eq.(\ref{r-tau-radial}), the radial equation of motion can also be obtained in terms of coordinate time $t$ (for $n=4$) as,
\begin{equation}
\frac{dt}{dr}=\pm \frac{1}{f}\left(1+\frac{p}{r}\right)^2,
\label{r-t-radial-null}
\end{equation}
where $f$ is given in eq.(\ref{eq:f_n_equal_q}). On integrating eq.(\ref{r-t-radial-null}), the radial coordinate $r$ can be obtained in terms of the coordinate time $t$. 
For the present study, the eq.(\ref{r-t-radial-null}) is solved numerically for an incoming test particle  (i.e. consideration of the negative sign in eq.(\ref{r-t-radial-null})).
The radial motion of a massless test particle with coordinate time and proper time is presented in fig.(\ref{time-r-null}).
One can observe from fig.(\ref{time-r-null}) that a massless test particle crosses the event horizon in a finite proper time while it approaches asymptotically to the event horizon in terms of coordinate time.
\newpage
\subsection{Non-radial Geodesics}
\begin{figure}[h!]
\centerline {\includegraphics[scale=0.36]{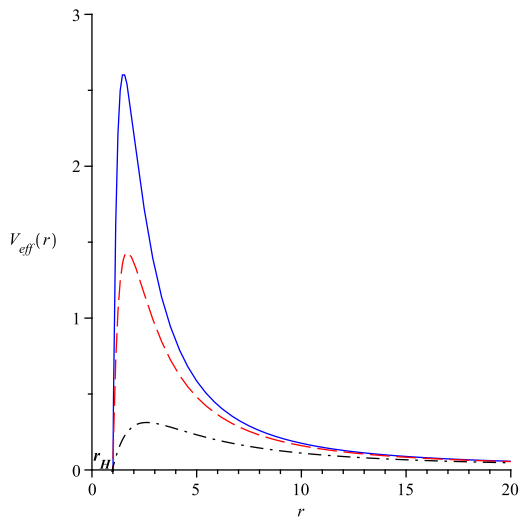}}
\vspace*{4pt}
\caption{\small{Effective potential for non-radial null geodesics with $\mu$ = 1, $L$ = 6, $g$ = 0.02, $n$ = 4, where for solid line $\beta$ = 0.1, for dashed line $\beta$ = 0.5, for dot-dashed line $\beta$ = 1, $r_H$ represents the position of event horizon.}}
\protect\label{Potential_Null_beta}
\end{figure}
\noindent In this section, we study the null geodesics for incoming test particles with non zero angular momentum.
Variation in the effective potential with charge parameter $\beta$ is shown in fig.(\ref{Potential_Null_beta}) while all other parameters are fixed.
One can observe from fig.(\ref{Potential_Null_beta}), that height of the effective potential decreases with increasing the value of charges for a particular value of $n$.\\
\begin{figure}[h!]
\centerline {\includegraphics[scale=0.4]{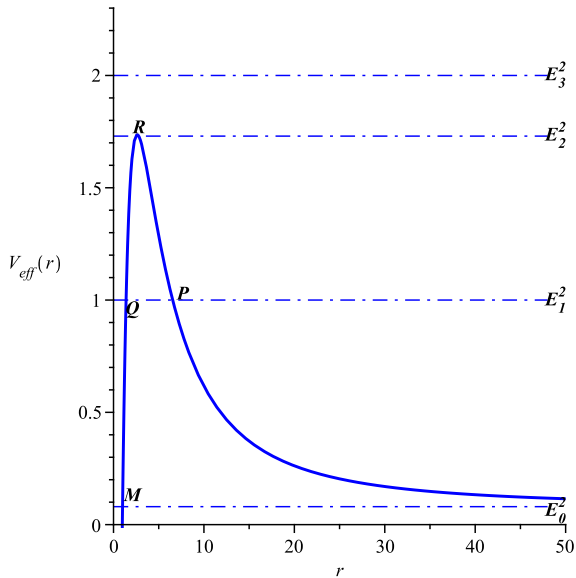}}
\vspace*{4pt}
\caption{\small{Effective potential of a unit mass black hole (i. e. $\mu$ = 1) for a massless test particle in the presence of four equal charges with $L$ = 10, $\beta$=1, $g$ = 0.02.}}
\protect\label{Potential_Null}
\end{figure}
\noindent 
The following orbits are allowed depending on the values of the constant $E$ (i.e. energy of the incoming test particle) as shown in fig.(\ref{Potential_Null}).\\
I) \underline{$E=E_0$:}\\
A particle starting form point $M$ drops into the singularity, forming a {\it terminating bound orbit}.\\
II) \underline{$E=E_1$:}\\
(i) A {\it fly-by orbit} is present for the particle coming from infinity, which turns at point $P$ as shown in fig.(\ref{Potential_Null}).\\
(ii) For the other possible orbit at this energy value, the test particle starts from the point $Q$ and falls into the singularity. Hence it is a {\it terminating bound orbit}.\\
III) \underline{$E=E_2$:}\\
A particle starting from infinity follows an {\it unstable circular orbit} at point $R$ which finally drops into the singularity as depicted in fig(\ref{Potential_Null}). Hence starting from point $R$, it follows a {\it terminating bound orbit}.\\
IV) \underline{$E=E_3$:}\\ 
A particle starting from the infinity finally drops into the singularity after crossing the horizon. Hence it is a {\it terminating escape orbit}.
\subsection{Orbit Analysis for null geodesics:}
\begin{figure}[h]
\centerline{ \includegraphics[scale=0.3]{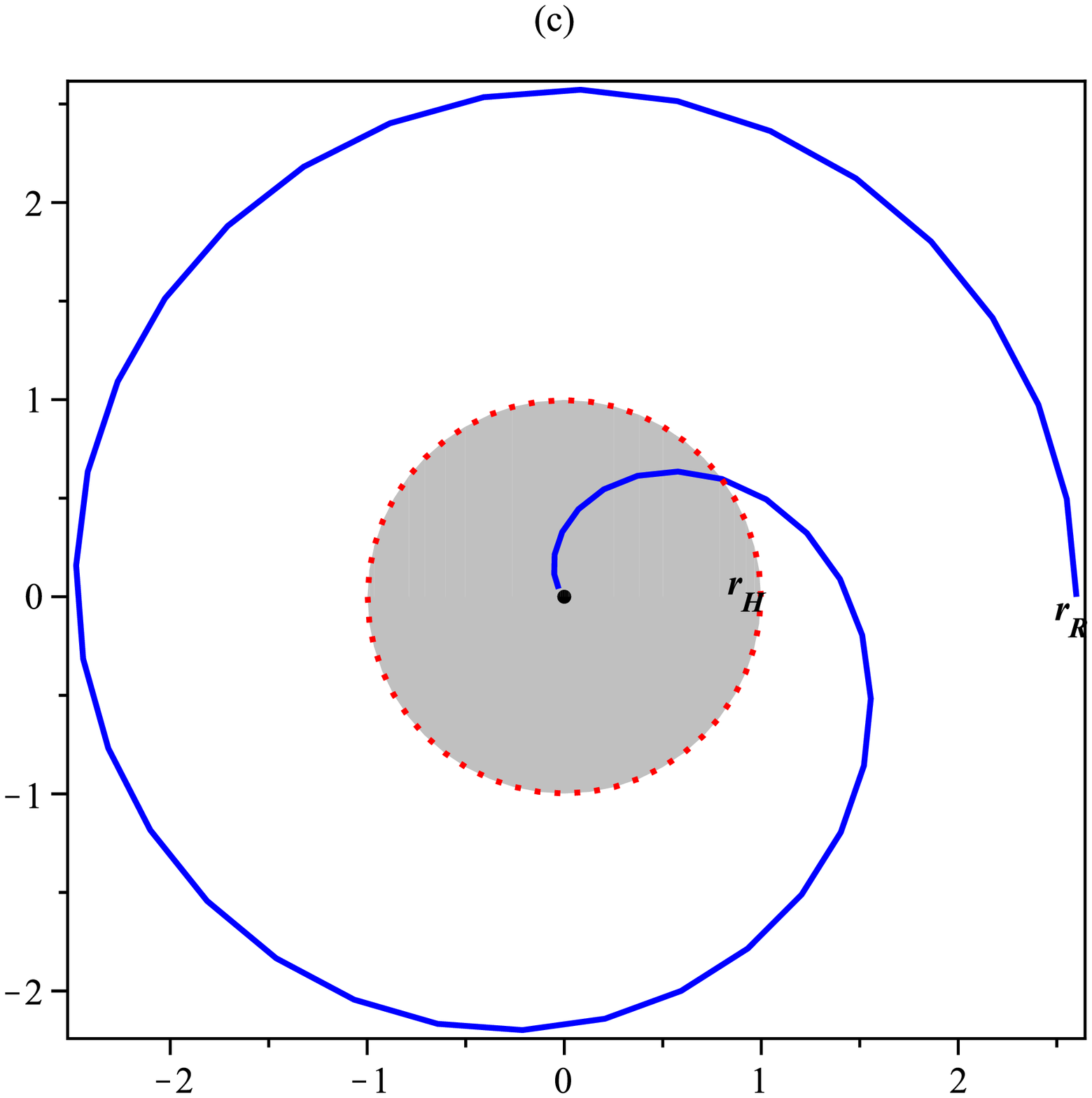}}
\vspace*{4pt}
\caption{\small{{\bf (a)} ${{P_4}^{null}}(r)$ represents the shaded allowed region for motion of a massless test particle one real positive zero at point $\bf{R}$,
{\bf (b)} Solid line represents the orbit of a massless test particle in region $II$,
{\bf (c)} Solid line represents the orbit of a massless test particle in region $I$ while dotted line represents the event horizon; with $L=10$, $\beta$=1, $g=0.02$, $E^2=1.736$, $n=4$.}}
\protect\label{N_E_C}
\end{figure}

\noindent Using the constraint for null geodesics (i.e. $u_{\mu}u^{\mu}$=0), the corresponding orbit equation reads as,
\begin{equation}
\left(\frac{dr}{d\phi}\right)^2=\frac{1}{L^2}{{P_n}^{null}}(r),
\label{r-phi-null}
\end{equation}
here
\begin{equation}
{{P_n}^{null}}(r)=\left(E^2 H^n-\frac{L^2 f}{r^2}\right)r^4.
\end{equation}
where, $f$ and $H$ are given in eq.(\ref{eq:f_n_equal_q}).
Again the physically acceptable regions having positive values of ${{P_n}^{null}}(r)$ or equivalently $E^2\ge{V_{eff}}$ for positive and real values of $r$, represents the allowed region of motion for massless test particles. The number of positive real zeros of ${{P_n}^{null}}(r)$ uniquely determine the type of orbits in the given background.
Let us now analyse all the possible types of orbits for massless test particles.
Hence, for $n=4$ and $\mu=1$ the polynomial ${{P_n}^{null}}(r)$ reduces to,
\begin{equation}
{{P_4}^{null}}(r)=E^2\left(r+{p}\right)^4-L^2\left[r^2-r+2g^2\left(r+{p}\right)^4\right].
\end{equation}
$\bf(I)$ \underline{\bf For E = $E_0$:}\\
There exists an interesting type of geodesics for massless test particles with definite angular momentum, specifically known as Cardioid type Geodesics \citep{Cruz2005}.
From eq.(\ref{r-phi-null}), it follows that with specific energy value of $E^2=2g^2L^2$, there also exist cardioid type geodesics for R-charged black holes.
For this special value of energy, the eq.(\ref{r-phi-null}) reduces to,
\begin{equation}
\left(\frac{dr}{d\phi}\right)^2=r(\mu-r).
\label{r-phi-cardioid}
\end{equation}
An elemental integration of eq.(\ref{r-phi-cardioid}) leads to,
\begin{equation}
r(\phi)=\frac{\mu}{2}(1+\cos\phi).
\end{equation}
which is the equation of the cardioid. This solution is independent of the angular momentum of incoming test particle and depends only on the mass of black hole.\\
$\bf(II)$ \underline{\bf For E = $E_1$:}\\
\begin{figure}[h]
\centerline{\includegraphics[scale=0.38]{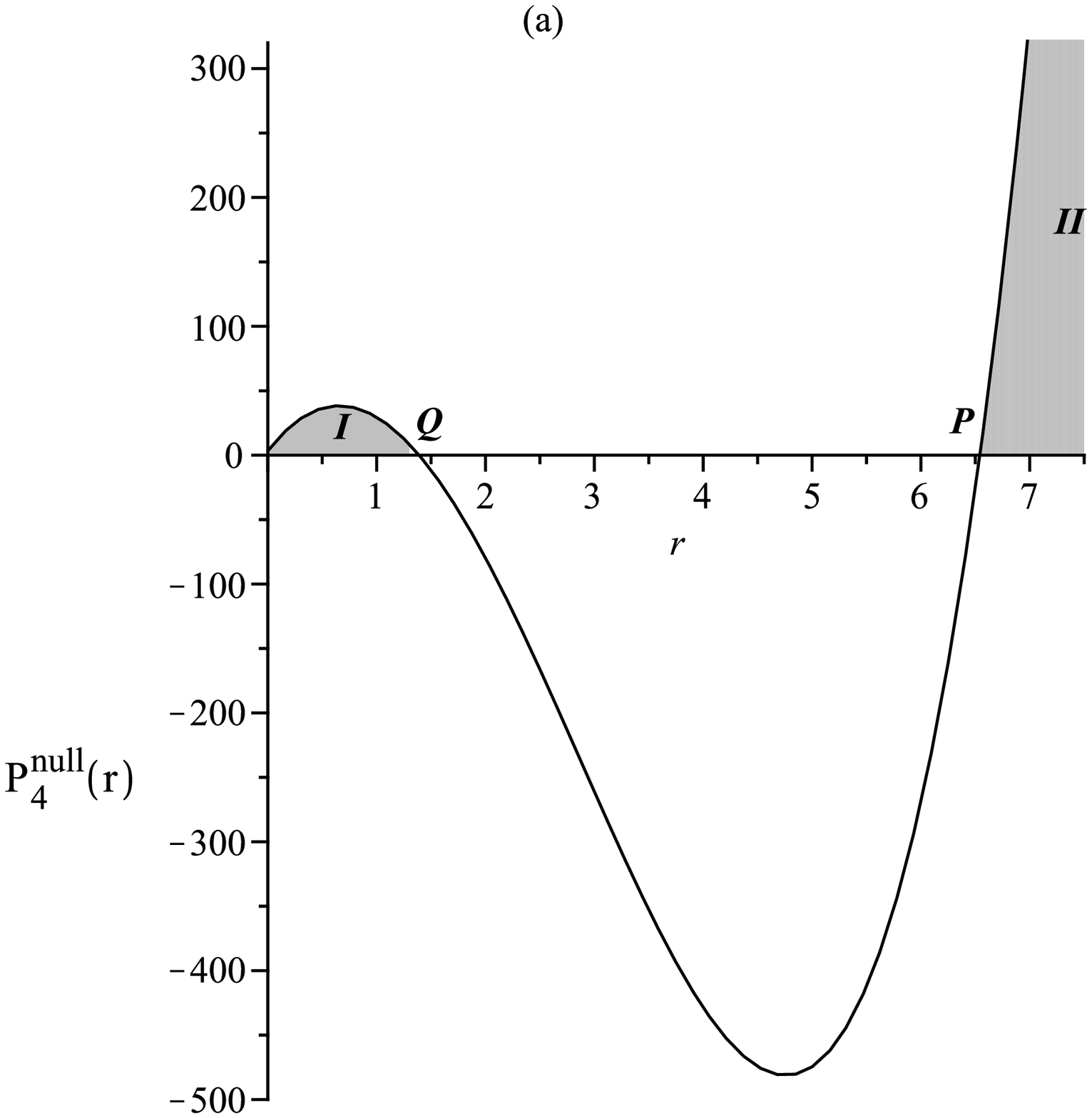}
\hskip5mm \includegraphics[scale=0.38]{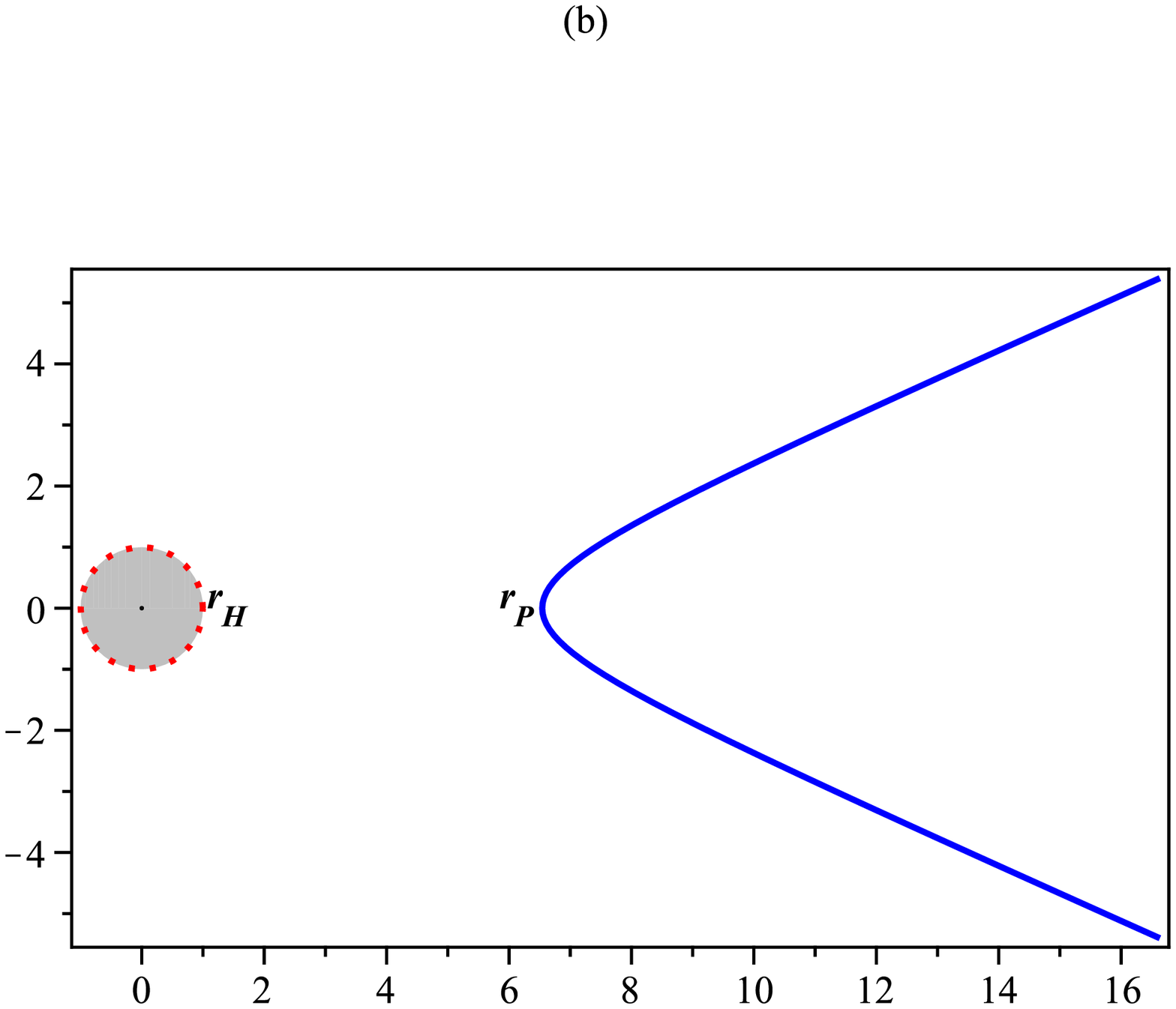}}
\nonumber
\end{figure}
\begin{figure}[h]
\centerline{ \includegraphics[scale=0.3]{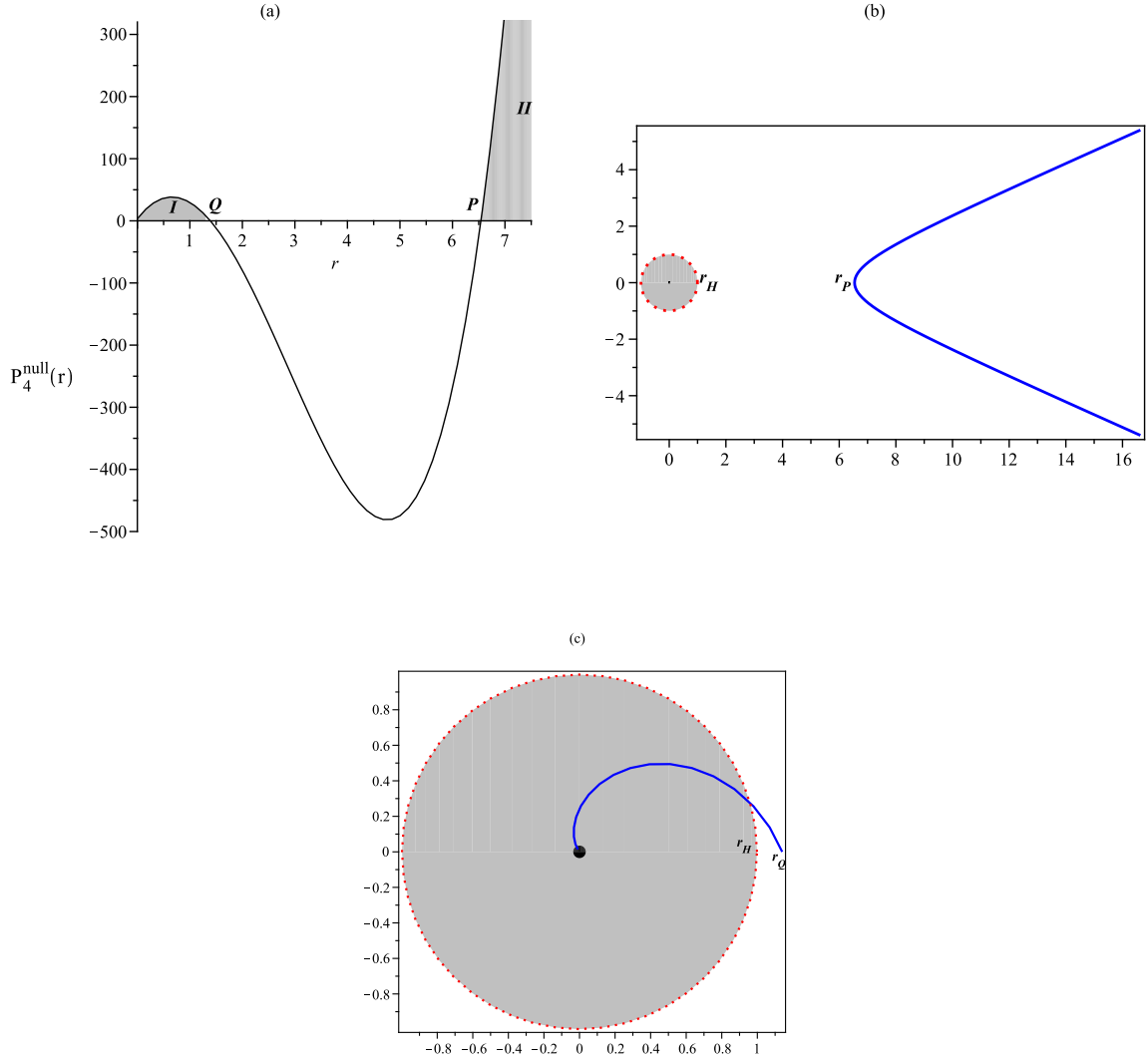}}
\vspace*{4pt}
\caption{\small{{\bf (a)} ${{P_4}^{null}}(r)$ represents the shaded allowed region for motion of a massless test particle with two real positive zeros at point $\bf{P}$ and $\bf{Q}$,
{\bf (b)} Solid line represents the orbit of a massless test particle in region $II$,
{\bf (c)} Solid line represents the orbit of a massless test particle in region $I$ while dotted line represents the event horizon; with $L=10$, $\beta$=1, $g=0.02$, $E^2=1$, $n=4$.}}
\protect\label{N_E_1}
\end{figure}
The numerical solution of orbit equation (\ref{r-phi-null}) is shown in fig.(\ref{N_E_1}).
Here one may notice the presence of a {\it fly-by orbit} with turning point at $P$ and a {\it terminating bound orbit} starting from point $Q$ as depicted in fig.(\ref{Potential_Null}).\\
\newpage
$\bf(III)$ \underline{\bf For E = $E_2$:}\\
The trajectory of a massless test particle with energy $E=E_2$ is presented numerically in fig.(\ref{N_E_C}). The particle starting from infinity forms an unstable circular orbit at point $R$, which finally terminates into the singularity.
\begin{figure}[h!]
\centerline{\includegraphics[scale=0.35]{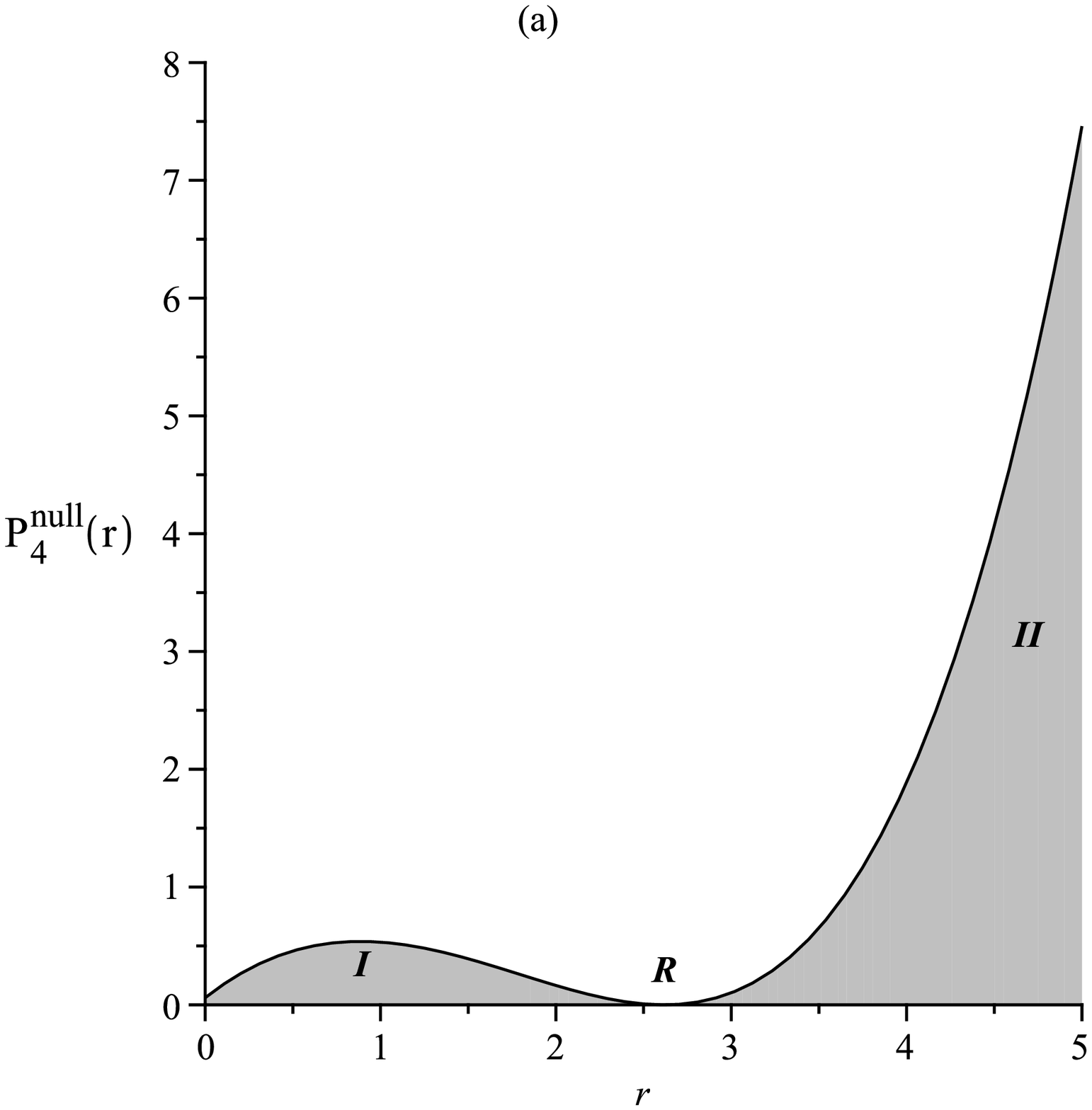}
\hskip5mm \includegraphics[scale=0.35]{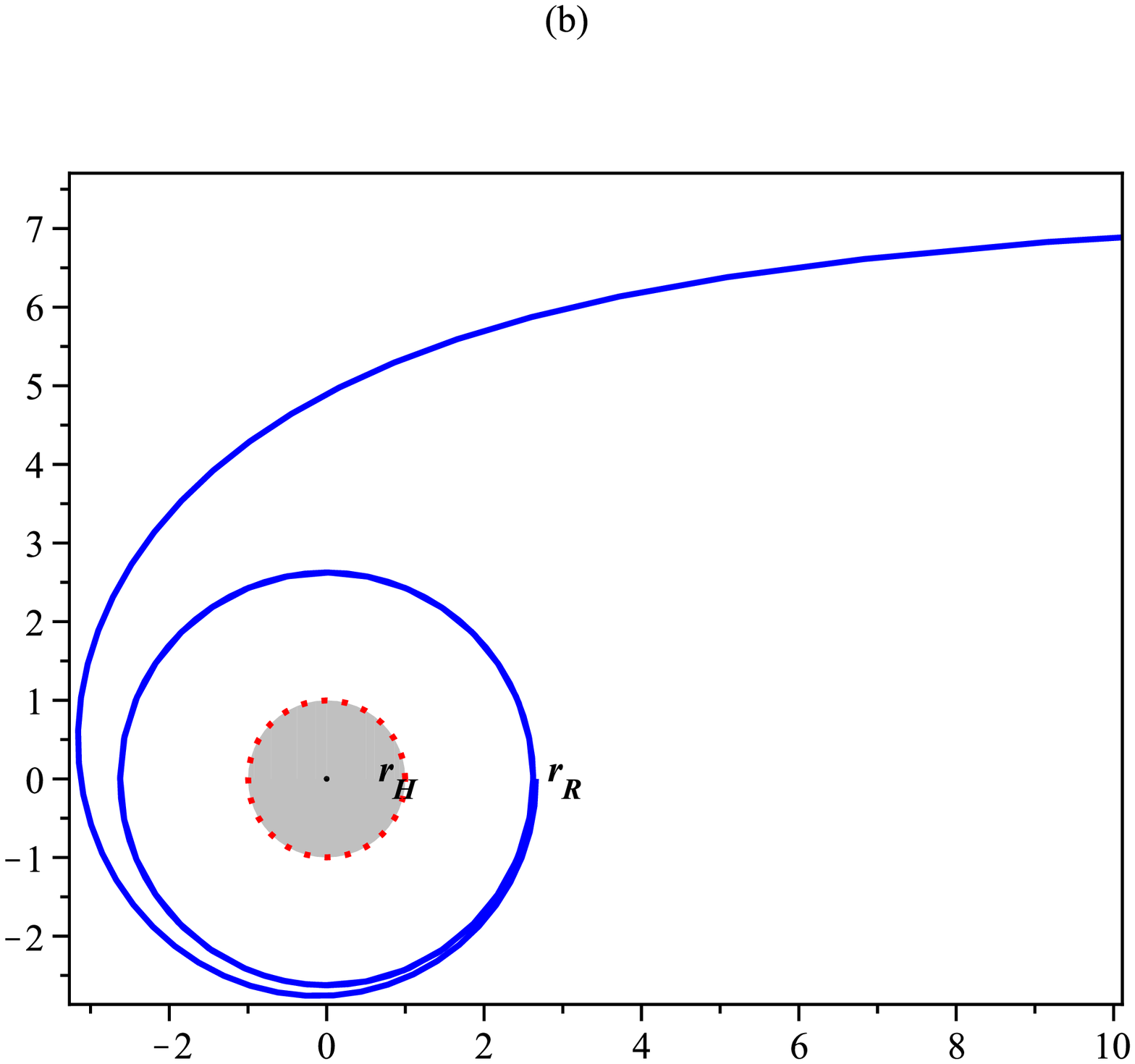}}
\nonumber
\end{figure}
\\
The circular orbit condition (i.e. ${V}^\prime_{eff}$(r) = 0) for null geodesics is reduces to,
\begin{equation}
2r^2-(2p+3\mu)r+\mu p=0,
\label{null-circular-condition}
\end{equation}
where $V_{eff}$(r) is given in eq.(\ref{eq:Potential_Non_Radial_Null}) with $n$ =4, $p$ = $\mu\sinh^2\beta$.
Solution of eq.(\ref{null-circular-condition}) describes the radius of {\it unstable circular orbit} and is given as,
\begin{equation}
r_c=\frac{1}{4}\left[2p+3\mu\pm\sqrt{\left(2p+3\mu\right)^2-8{\mu}p}\right],
\label{r-circular-null}
\end{equation}
where $p=\mu\sinh^2\beta$ and eq.(\ref{null-circular-condition}) has a real solution for $\left(2p+3\mu\right)^2\geq8{\mu}p$. The larger root of eq.(\ref{null-circular-condition}) locates the position of unstable circular orbit, which again has the minimum value for the condition $\left(2p+3\mu\right)^2=8{\mu}p$ as,
\begin{equation}
{(r_c)_{min}}=\frac{1}{4}\left(2p+3\mu\right).
\label{r-circle-min}
\end{equation}
It is clear from eq.(\ref{r-circular-null}) that radius of unstable circular orbit does not depend on the coupling constant $g$.
In the absence of charges (i.e. $p\rightarrow0$), eq.(\ref{r-circular-null}) and eq.(\ref{r-circle-min}) both reduces to the Schwarzschild case. Hence due to the presence of charges, the radius of unstable circular orbits is increased.\\
{\it\underline {The Time Period:}}
The time period for unstable circular orbits can be calculated for proper time as well as coordinate time.
The expression for time period for proper time can be obtained by integrating eq.(\ref{Com2}) (with $\phi$=2$\pi$ for one time period) as,
\begin{equation}
T_{\tau}= \frac{2\pi {r_{_C}}^2}{L}H^{n/2}(r_{_C}).
\label{time-period-proper}
\end{equation}
Time period in coordinate time can be obtained by combining eq.(\ref{Com1}) and eq.(\ref{Com2}) and integrating afterwards as follows,
\begin{equation}
T_t=2\pi r_{_C}\left(\frac{{H^n}(r_{_C})}{f(r_{_C})}\right)^{1/2}.
\label{time-period-coordinate}
\end{equation}
\begin{table}[h!]
\centering
\begin{tabular}{ | c | c | c | c | }
\hline
$n$ & $r_{_C}$ & $T_t$ & $T_\tau$ 
\\
\hline
\hline
0 & 1.5000 & 5.1822$\pi$ & 6.2366$\pi$ \\
\hline
1 & 1.6477 & 7.0904$\pi$ & 7.3617$\pi$  \\
\hline
2 & 1.8694 & 9.4472$\pi$ & 12.153$\pi$  \\
\hline
3 & 2.1906 & 12.188$\pi$ & 19.982$\pi$ \\
\hline
4 & 2.6174 & 15.179$\pi$ &  31.979$\pi$   \\
\hline
\end{tabular} 
\caption{Comparison of the time periods for unstable circular orbits of massless test particles with different number of charges (i.e. $n$) for $\mu=1$, 
$g=0.02$ and $\beta=1$.}
\label{table:2}
\end{table}
\newpage
\noindent{\it \underline{Cone of Avoidance:}}\\
 The cone of avoidance can be defined at any point \citep{Fernando2012,Cruz2005,Schandra1992} and the light rays included in the cone must necessarily cross the horizon and get trapped.
If $\psi$ denotes the half-angle of the cone directed towards the black hole at large distances, then
\begin{equation}
\cot\psi=+\frac{1}{\mathcal{R}}\frac{d\tilde{r}}{d\phi},
\end{equation}
here
\begin{equation}
d\tilde{r}=\frac{(r+p)}{rf^{1/2}}dr,
\label{coa-proper-length}
\end{equation}
where $f$ is given by eq.(\ref{eq:f_n_equal_q}). The radial function $\mathcal{R}$ described in eq.(\ref{dA_prime}) and eq.(\ref{R_approx}) is given as $\mathcal{R}= (r+p)$ for $n=4$. Eq.(\ref{coa-proper-length}) describes the element of proper length along the generator of the cone. Hence,
\begin{equation}
\cot\psi=\frac{1}{rf^{1/2}}\frac{dr}{d\phi}.
\end{equation}
Now using the eq.(\ref{r-phi-null}), one can obtain
\begin{equation}
\tan\psi=\left[\frac{L^2fr^2}{E^2(r+p)^4-L^2fr^2}\right]^{1/2}.
\label{coa-tanpsi}
\end{equation}
From eq.(\ref{coa-tanpsi}), it follows that,
\begin{equation}
\psi=\frac{\pi}{2}\hspace{2mm} for \hspace{2mm} r=r_c,
\nonumber
\end{equation}
\begin{equation}
\psi=0\hspace{2mm} for \hspace{2mm} r=r_H,
\nonumber
\end{equation}
\begin{equation}
\psi\approx\sqrt{\frac{1+6p^2+2g^2(4rp+r^2)}{\left(\frac{E^2}{L^2}-2g^2\right)(6p^2+4rp+r^2)}} \hspace{2mm} for \hspace{2mm} r>>1,
\end{equation}
where, $r_c$ denotes the radius of circular orbit and $r_H$ denotes the radius of event horizon. 
\newpage
$\bf(IV)$ \underline{\bf For E = $E_3$:}\\
\begin{figure}[h!]
\centerline{\includegraphics[scale=0.35]{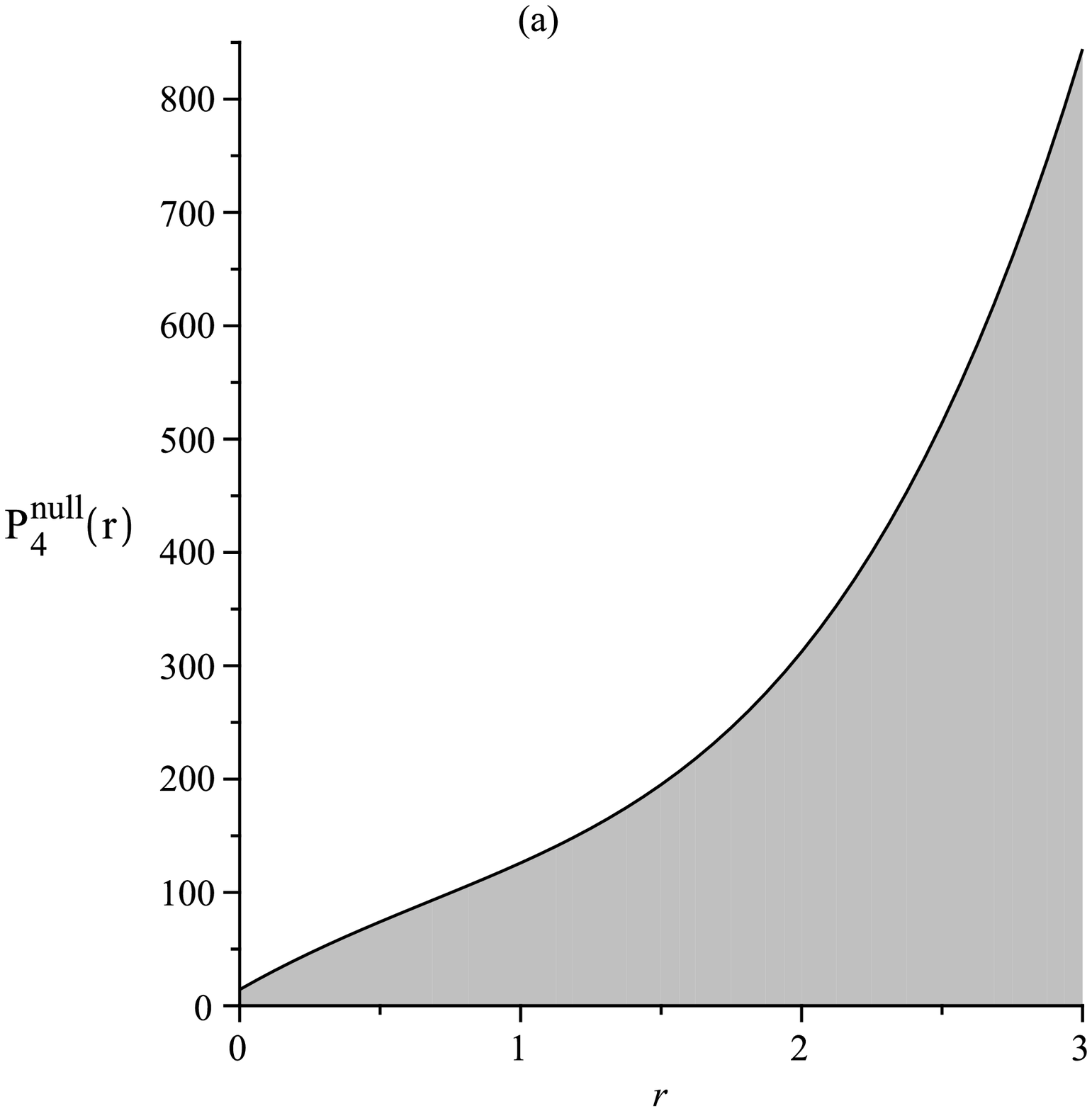}
\hskip5mm \includegraphics[scale=0.35]{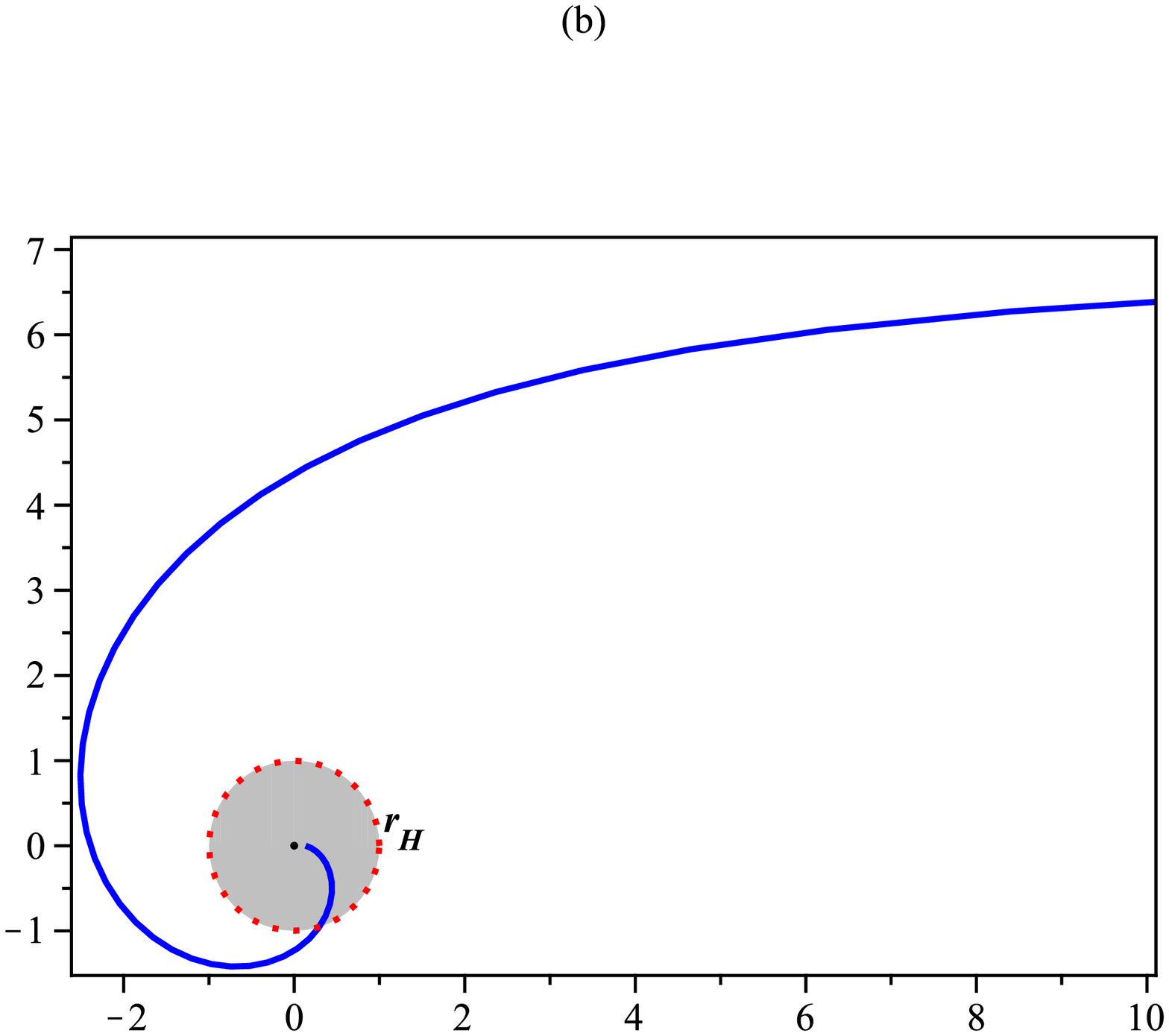}}
\vspace*{4pt}
\caption{\small{{\bf (a)} ${{P_4}^{null}}(r)$ represents the shaded allowed region for motion of massless test particle no real positive zero,
{\bf (b)} Solid line represents the orbit of massless test particle in the allowed region while dotted line represents the event horizon; with $L=10$, $\beta$=1, $g=0.02$, $E^2=2$, $n=4$.}}
\protect\label{N_E_3}
\nonumber
\end{figure}
The path of a massless test particle with energy $E=E_3$ is simulated numerically in fig.(\ref{N_E_3}) which shows the presence of a terminating escape orbit.\\

\section{Summary and Conclusions}
\noindent In this article, we have investigated geodesic motion of massive as well as massless test particles around a particular class of R-charged black holes in $N=8$, $D=4$ gauged supergravity theory. 
Some of the important results are summarised below:
\begin{itemize}
\item It is observed that  in R-charged black hole spacetimes, the horizon is always present for single and two charges, while it occurs
only for a particular condition in case of charges more than two.

\item Different types of orbits (such as terminating bound, planetary, stable and unstable circular orbits) are present for incoming massive test particles, corresponding to their initial energy.
No fly-by orbits are observed in any case for massive test particles.
Fly-by and terminating escape orbits are however present for the case of massless test particles unlike the case of massive test particles.
There exist no stable circular orbits for a massless test particle.
An interesting type of orbit for a massless test particle namely, Cardioid type geodesic is found to depend only on the mass of the black hole.

\item Radius of stable circular orbits for massive test particles and radius of unstable circular orbits for massless test particles are found to increase with the enhancement in the number of charges. We have also computed the time periods both  in coordinate and proper time to visualise the changes with the number of charges.
The radius of unstable circular orbit for a massless test particle is found to be independent of the gauge coupling constant.

\item Advances of perihelion in planetary orbits is calculated for massive test particles by using the Keplerian orbit method and  the corrections arising from non zero charge parameters apart from the gauge coupling constant are obtained.

\item 
Cone of avoidance for a massless test particle is calculated and it is observed that at large distances, it depends on both the charge and gauge coupling constant.\\
\vspace{1mm}\\
\noindent All the results obtained in the present study reproduces the results manifestly corresponding to well known spacetimes such as Schwarzschild and Schwarzschild AdS black hole spacetimes in the prescribed limits respectively. 
\end{itemize}
\section*{Acknowledgments}
\noindent 
The authors are indebted to the anonymous referee for his useful suggestions and comments which helped us to improve the presentation of the manuscript significantly.
One of the authors HN would like to thank Department of Science and Technology, New Delhi for financial support through grant no. SR/FTP/PS-31/2009.
HN and RU acknowledge the  support from CTS, IIT, Kharagpur under its visitors program. 
RU also acknowledges the  support from IUCAA, Pune under its visitors program.


\begin{references}
\bibitem{Eins} A. Einstein, Sitzung der physikalisch-mathematischenklasse \textbf{25}, 844 (1915).
\bibitem{Poi} E. Poisson, \textit{A relativists’ toolkit: the mathematics of black hole mechanics}(Cambridge University Press, 2004).
\bibitem{Wald} R. M. Wald, \textit{General Relativity} (University of Chicago Press, Chicago, USA, 1984)
\bibitem{Joshi} P. S. Joshi, \textit{Global aspects in gravitation and cosmology} (Oxford University Press, Oxford, UK, 1997).
\bibitem{Schwar} K. Schwarzschild, Sitzungsber. Preuss. Akad. Wiss. Berlin (Math. Phys. Tech.), 189 (1916).
\bibitem{Sen2005} A. Sen, Curr. Sci. \textbf{88}, 2045 (2005).
\bibitem{Zwie2004} B. Zwiebach, \textit{A first course in string theory} (Cambridge University Press, 2004).
\bibitem{Pol1998} J. Polchinsk, \textit{String theory Vol. 1 and 2} (Cambridge University Press, 1998).
\bibitem{Wit1982}  B. de. Wit and H. Nicolai, Phys. Lett.\textbf{B108}, 285 (1982).
\bibitem{Nic1982} B. de. Wit and H. Nicolai, Nucl. Phy.\textbf{B208}, 323 (1982).
\bibitem{Duff1986} M. J. Duff and B. E. W. Nillson and C. N. Pope, Phys. Rept. \textbf{130}, 1 (1986).
\bibitem{Malda1998} J. M. Maldacena, Adv. Theor. Math. Phys \textbf{2}, 231 (1998), hep-th/9711200. 
\bibitem{Witt1998} E. Witten, Adv. Theor. Math. Phys. \textbf{2}, 253 (1998), hep-th/9802150.
\bibitem{Witten1998} E. Witten, Adv. Theor. Math. Phys. \textbf{2}, 505 (1998), hep-th/9803131.
\bibitem{Liu1999}  M. J. Duff and James. T. Liu, Nucl. Phy.\textbf{B554}, 237 (1999), hep-th/9901149v2.
\bibitem{Cvetic1999} M. Cvetic and S. S. Gubser, JHEP \textbf{9904}, 024 (1999), hep-th/9902195.
\bibitem{Duf1999} M. J. Duff,  Int. J. Mod. Phys. \textbf{A14} 815 (1995), hep-th/9808100.
\bibitem{Hagi1931} Y. Hagihara, Jpn. J. Astron. Geophys. \textbf{8} 67, (1931).
\bibitem{Kagra2010} V. Kagramanova and J. Kunz and E. Hackmann and C. Lammerzahl, Phys. Rev \textbf{D81} 124044 (2010), arXiv:1002.4342 [gr-qc].
\bibitem{Grunau2011} S. Grunau and V. Kagramanova, Phys. Rev \textbf{D83} 044009 (2011), arXiv:1011.5399 [gr-qc].
\bibitem{Kagra2011} V. Kagramanova and S. Reimers, Phys. Rev \textbf{D86} 084029 (2011), arXiv:1208.3686 [gr-qc].
\bibitem{Heck2008} E. Hackmann and C. Lammerzah, Phys. Rev. Lett. \textbf{100} 171101 (2008), arXiv:1505.07955 [gr-qc].
\bibitem{Hec2008}  E. Hackmann and C. Lammerzah, Phys. Rev. \textbf{D78} 024035 (2008), arXiv:1505.07973 [gr-qc].
\bibitem{Heckm2008} E. Hackmann and V. Kagramanova and J. Kunz and C. Lammerzahl, Phys. Rev. \textbf{D78} 124018 (2008), arXiv:0812.2428 [gr-qc].
\bibitem{Heck2010}  E. Hackmann and C. Lammerzahl and V. Kagramanova and J. Kunz, Phys. Rev. \textbf{D81} 044020 (2010), arXiv:arXiv:1009.6117 [gr-qc].
\bibitem{Enol2011} V. Z. Enolski and E. Hackmann and V. Kagramanova and J. Kunz and C. Lammerzahl, J. Geom. Phys. \textbf{61} 899 (2011), arXiv:1011.6459 [gr-qc].
\bibitem{Grunau2012} S. Grunau and V. Kagramanova and J. Kunz and C. Lammerzah, Phys. Rev. \textbf{D86} 104002 (2012), arXiv:arXiv:1208.2548 [gr-qc].
\bibitem{Grunau2013}  S. Grunau and V. Kagramanova and J. Kunz, Phys. Rev. \textbf{D87} 044054 (2013), arXiv:arXiv:1212.0416 [gr-qc].
\bibitem{Oliv2011} M. Olivares and J. Saavedra and C. Leiva and J. R. Villanueva, Mod. Phys. Lett. \textbf{A26} 2923 (2011), arXiv:1101.0748 [gr-qc].
\bibitem{Vill2013} J. R. Villanueva and J. Saavedra and M. Olivares and N. Cruz, Astrophys. Space Sci. \textbf{344}, 437 (2013). 	
\bibitem{Sor2015} S. Soroushfar and R. Saffari1 and J. Kunz and C. Lammerzah, (2015), arXiv:1504.07854.
\bibitem{Uniyal2015} R. Uniyal and N. C. Devi and H. Nandan and K.D. Purohit, Gen.Rel.Grav. \textbf{47}, 16 (2015), arXiv:1406.3931[gr-qc].
\bibitem{Ghosh2010} S. Ghosh and S. Kar and H. Nandan, Phys. Rev. \textbf{D82}, 024040 (2010), arXiv:0904.2321[gr-qc].
\bibitem{Nandan2010} H. Nandan and N. M. Bezares-Roder and H. Dehnen, Class. Quant. Grav \textbf{27}, 245003 (2010), arXiv:0912.4036[gr-qc].
\bibitem{Das2012} A. Dasgupta and H. Nandan and S. Kar, Phys. Rev. \textbf{D85}, 104037 (2012), arXiv:1202.5370[gr-qc].
\bibitem{Koley2003} R. Koley and S. Pal and S. Kar, Am. J. Phys. \textbf{71}, 1037 (2003).
\bibitem{Kuniyal2015} R. S. Kuniyal and R. Uniyal and H. Nandan and A. Zaidi, Astrophys. Space Sci. \textbf{357}, 92 (2015), arXiv:1505.00103[gr-qc].
\bibitem{Uniyal2014} R. Uniyal and H. Nandan and K.D. Purohit, Mod.Phys.Lett. \textbf{A29}, 1450157 (2014), arXiv:1406.3918[gr-qc].
\bibitem{Das2009} A. Dasgupta and H. Nandan and S. Kar, Phys. Rev. \textbf{D79}, 124004 (2009), arXiv:0809.3074[gr-qc].
\bibitem{Gad2010} R. M. Gad, Astrophys. Space Sci. \textbf{330}, 107 (2010), arXiv:0708.2841[gr-qc].
\bibitem{Fernando2012} S. Fernando, Phys. Rev. \textbf{D85}, 024033 (2012), arXiv:1109.0254[hep-th].
\bibitem{Oliva2013} M. Olivares, Eur.Phys.J. \textbf{C73}, 2659 (2013), arXiv:1311.4236[gr-qc].
\bibitem{Saskia2013} S. Grunau and B. Khamesra, Phys. Rev. \textbf{D87}, 124019 (2013), arXiv:1303.6863[gr-qc].
\bibitem{Romans1992} L. J. Romans, Nucl.  Phys. \textbf{B383}, 395 (1992), arXiv:hep-th/9203018.
\bibitem{Corn1993} S. Cornbleet, Am. J.  Phys. \textbf{61}, 7 (1993).
\bibitem{Cruz2005} N. Cruz and M. Olivares and J. R. Villanueva,  Class. Quant. Grav. \textbf{22}, 1167 (2005), arXiv:gr-qc/0408016.
\bibitem{Schandra1992} S. Chandrashekha, \textit{The Mathematical Theory of Black Holes  646p} (Oxford, UK, 1992).
\end{references}
\end{document}